\documentclass[a4paper]{article}
\usepackage[left=2.5cm,right=2.5cm,top=2cm,bottom=2cm]{geometry}
\usepackage{url} 
\usepackage{natbib}
\usepackage{caption}
\usepackage{subcaption}
\usepackage{amsmath}
\usepackage{amsfonts}
\usepackage{graphicx}
\usepackage{wrapfig}
\usepackage{lscape}
\usepackage{rotating}
\usepackage{multirow}
\usepackage{float}
\usepackage{authblk}
\usepackage{xcolor}

\begin{document}

\title{\hrule

\vspace{10pt}

\Large \textbf{Explainable Artificial Intelligence for Bayesian Neural Networks: Towards  trustworthy predictions of ocean dynamics}

\vspace{10pt}

\hrule}

\author[1]{\large Mariana C. A. Clare}
\author[2,3,4]{\large Maike Sonnewald}
\author[5]{\large Redouane Lguensat}
\author[6]{\large Julie Deshayes}
\author[2,3,7]{\large V. Balaji}

\affil[1]{\footnotesize Imperial College London, London, UK}
\affil[2]{Princeton University, Program in Atmospheric and Oceanic Sciences, Princeton, USA}
\affil[3]{NOAA/Geophysical Fluid Dynamics Laboratory, Ocean and Cryosphere Division, Princeton, USA}
\affil[4]{University of Washington, Seattle, Washington, USA}
\affil[5]{Institut Pierre-Simon Laplace, IRD, Sorbonne Universit\'e, Paris, France}
\affil[6]{LOCEAN-IPSL, CNRS, Sorbonne Universit\'e, Paris, France}
\affil[7]{Laboratoire des Sciences du Climat et de l'Environnement, CEA Saclay, Gif Sur Yvette, France}

\date{}
\maketitle

\begin{abstract}
The trustworthiness of neural networks is often challenged because they lack the ability to express uncertainty and explain their skill. This can be problematic given the increasing use of neural networks in high stakes decision-making such as in climate change applications. We address both issues by successfully implementing a Bayesian Neural Network (BNN), where parameters are distributions rather than deterministic, and applying novel implementations of explainable AI (XAI) techniques. The uncertainty analysis from the BNN provides a comprehensive overview of the prediction more suited to practitioners' needs than predictions from a classical neural network. Using a BNN means we can calculate the entropy (\textit{i.e.} uncertainty) of the predictions and determine if the probability of an outcome is statistically significant. To enhance trustworthiness, we also spatially apply the two XAI techniques of Layer-wise Relevance Propagation (LRP) and SHapley Additive exPlanation (SHAP) values. These XAI methods reveal the extent to which the BNN is suitable and/or trustworthy. Using two techniques gives a more holistic view of BNN skill and its uncertainty, as LRP considers neural network parameters, whereas SHAP considers changes to outputs. We verify these techniques using comparison with intuition from physical theory. The differences in explanation identify potential areas where new physical theory guided studies are needed.
\end{abstract}

\section{Introduction}\label{sec:intro}
There is already scientific certainty that global heating is changing the climate, but understanding exactly how the climate will change and the potential impacts is an open problem. Increasingly, artificial intelligence techniques, such as neural networks, are being used to better understand climate change \citep[for example][]{ham2019deep,huntingford2019machine,rolnick2019tackling,cowls2021ai}, but as neural network techniques become evermore ubiquitous, there is a growing need for methods to quantify their trustworthiness and uncertainty \citep{li2021trustworthy,mamalakis2021neural}. Following \cite{THOR}, we define a method to be trustworthy if its results are explainable and interpretable, and therefore these two concepts are somewhat linked as improving uncertainty quantification also improves result interpretability. Quantifying uncertainty using classical neural networks is particularly difficult because they lack the ability to express it and are often overconfident in their results \citep{mitros2019validity,Joo2020}. A range of techniques have been used to address this uncertainty quantification issue \citep{guo2017calibration} and a particularly common one is to use an ensemble of deep learning models \citep[for example][]{beluch2018power}. However, choosing a good ensemble of models is non-trivial \citep[see][]{Scher2020} and may be computationally expensive because it requires the network to be trained multiple times. This lack of uncertainty analysis limits the extent to which classical neural networks can be useful for ocean and climate science problems. For example, lack of knowledge of uncertainties in future projections of sea level rise limits how effective coastal protection measures can be for coastal communities \citep{sanchez2021coastal}. Measures of uncertainty are also important for out-of-sample predictions, which are common in climate change science because neural networks must be trained on historical data and applied to a changed climate scenario where the dynamics governing a region may have fundamentally changed. Thus, quantifying uncertainty within a climate application is of paramount importance as decisions based on neural network predictions could have wide ranging impacts. Moreover, there can be distrust of neural network predictions in the climate science community because of the potential for spurious correlations giving rise to predictions that are nonphysical. Predictions are more trustworthy if they are explainable (\textit{i.e.} if the reason why the network predicted the result can be understood by members of the climate science community). However, adding explainability techniques to uncertainty analysis is an understudied area.

In this work, we address both issues of uncertainty and trustworthiness by implementing a Bayesian Neural Network (BNN) \citep{jospin2020hands} with novel implementations of explainable AI techniques (known as XAI) \citep{samek2021explaining}. We focus on applying this technique to assess uncertainty in dynamical ocean regime predictions due to a changing climate following the THOR (Tracking global Heating with Ocean Regimes) framework \citep{THOR}. This is the first time BNNs have been used to predict large-scale ocean circulations, although they have been used for localised streamflows in \cite{rasouli2012daily, rasouli2020forecast}. Our work is particularly pertinent with a recent IPCC Special Report \citep{hoegh2018impacts} highlighting uncertainty in ocean circulation as a key knowledge gap area that must be addressed. Both \citep{THOR} and our work are designed to predict future changes to ocean circulation using data from the sixth phase of the Coupled Model Intercomparison Project (CMIP) (used in IPCC reports) \citep{eyring2015overview}. We note however that, as CMIP6 is a large international collaboration, data dissemination and quality control can be difficult, which in turn limits the capability for good analysis. \cite{THOR} is an example of using sparse data in this context, and resolving this issue generally is an area of ongoing research \citep{eyring2019}.

Unlike classical neural networks, BNNs make well-calibrated uncertainty predictions \citep{mitros2019validity,jospin2020hands} and clearly inform the user of how unsure the outcome is. This provides a more comprehensive description of the neural network prediction compared to a classical neural network and one which better meets the needs of climate and ocean science researchers. Furthermore, the uncertainty measures provided by the BNN approach reveal whether a prediction made on a sample that differs greatly from the training data can be trusted. For example, it is known that the wind stress over the Southern Ocean will change in the future, with implications for the dynamics key to maintaining global scale heat transport. However, the region already has extreme conditions, so a change here could result in entirely new dynamical connections. The BNN outputs would allow us to understand if the prediction based on the new conditions can still be trusted. This uncertainty analysis is possible in BNNs because the weights, biases and/or outputs are distributions rather than deterministic point values. Moreover, these distributions mean BNNs can easily be used as part of an ensemble approach (a very common approach in climate science), by simply sampling point estimates from the trained distributions to generate an ensemble \citep{Bykov2020}.

Using BNNs is a large step towards trustworthy predictions, but results also gain considerable trustworthiness to climate researchers and practitioners if their skill is physically explainable. Note that throughout we define explaining skill to mean explaining the correlations between the input features that give rise to the predictions. Governments and regulatory bodies are also increasingly imposing regulations that require trustworthiness in AI processes used in certain decision-making \citep[see][]{cath2018artificial} and imposing large fines if the standards are not met (see for example recent directives from the \cite{EU_law} and the USA government (E.O. 13960 of Dec 3, 2020)). XAI techniques can be used to explain the skill of neural networks \citep{samek2019explainable,samek2021explaining,arrieta2020explainable}, but there has been little work combining explainability with uncertainty analysis in part because the distributions in BNNs add extra complexity. In this work, we adapt two common XAI techniques so that they can be used to explain the skill in BNN results: Layer-wise Relevance Propagation (LRP) \citep{binder2016layer} which is here applied to BNNs for only the second time after having been first applied to BNNs in \cite{Bykov2020} and SHAP values \citep{lundberg2017unified} which are here applied to a BNN for the first time. These XAI methods reveal the extent to which the BNN is fit for purpose for our problem. Moreover, our approach means we can gain a reliable notion of the confidence of the explanation, which has been highlighted as a key area where XAI techniques must improve \citep{lakkaraju2022rethinking}. Applying our XAI techniques to BNNs trained on real-world ocean circulation data in an application designed to understand future climate has the added benefit that we are able to validate and confirm these novel applications of XAI using physical understanding of ocean circulation processes, improving confidence in our BNN predictions. Thus, our novel framework is able to quantify uncertainty and improve trustworthiness (\textit{i.e.} explainability and interpretability) in predictions, marking a significant step forward for using neural networks in climate and ocean science.

In this work, we choose to apply two different XAI techniques specifically to gain a holistic view of the skill of the BNN as LRP considers the neural network parameters whereas SHAP considers the impact of changing input features on the BNN outputs. This is important to ensure that what the BNN has learned is genuinely rooted in physical theory. The two different approaches also give a more overall impression of uncertainty as they capture different aspects with LRP capturing model uncertainty and SHAP capturing prediction sensitivity to this model uncertainty. Furthermore, by considering two different techniques, we can explore whether they agree as to which features are important in each region of the domain. This allows us test if the `disagreement problem' exists in this work, where two techniques explain network skill in different ways \citep{krishna2022disagreement}, which is a growing area of interest in XAI research. 

To summarise the main contributions of our work are that we present the first application of BNNs to quantify uncertainty in large-scale ocean circulation predictions, and explain the skill of these predictions through novel implementations of the XAI techniques, SHAP and LRP, thereby improving trustworthiness. The remainder of this paper is structured as follows: Section \ref{sec:methods} explores the theory behind BNNs and applying XAI techniques to BNNs, Section \ref{sec:data} explores the dataset used to train the BNN, Section \ref{sec:results} shows the results of applying the BNN and novel XAI techniques to the dataset and finally Section \ref{sec:conclusion} concludes this work.

\section{Methods}\label{sec:methods}
\subsection{Bayesian Neural Networks (BNNs)}\label{sec:BNN}
\begin{figure}
\begin{subfigure}{0.47\textwidth}
    \centering
    \includegraphics[width=\textwidth]{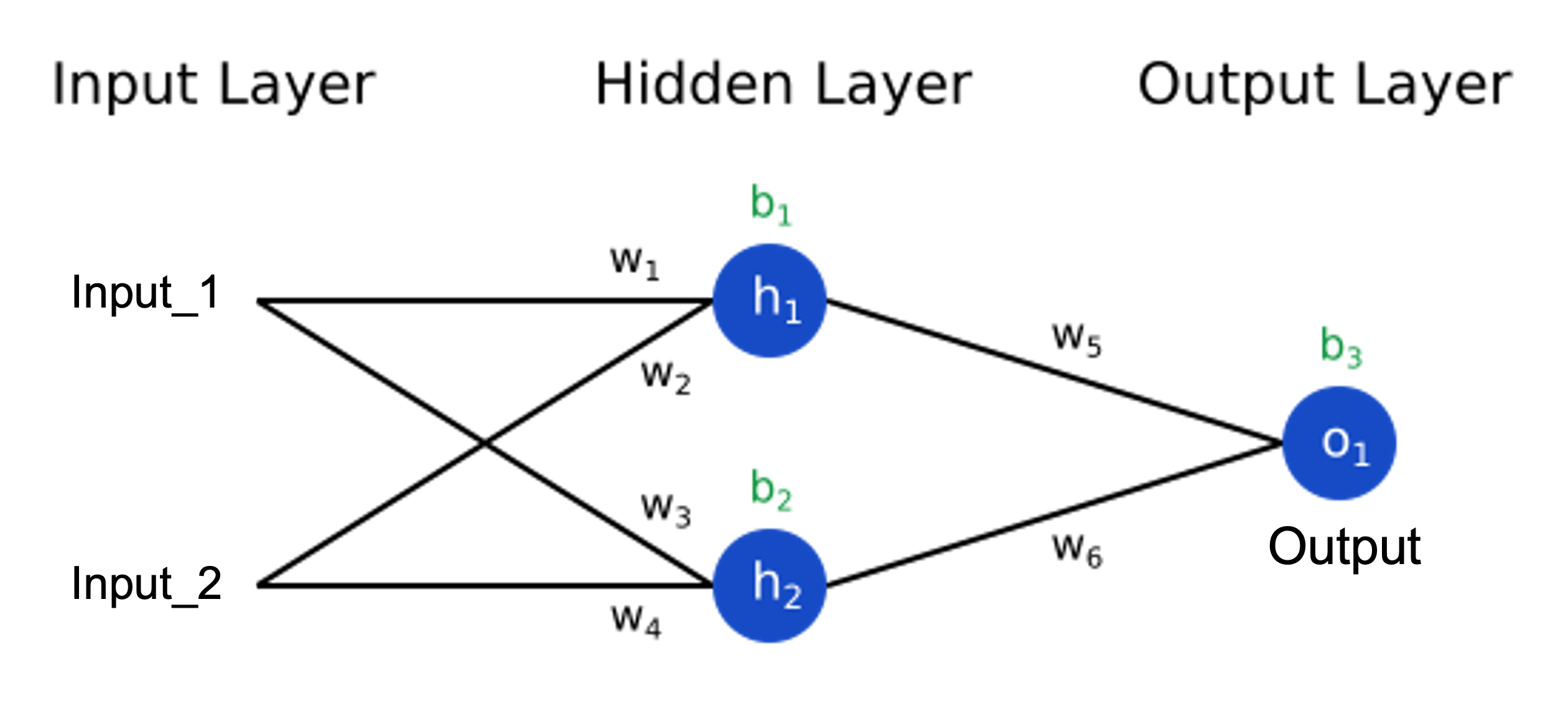}
    \caption{Classical deterministic Neural Network. Weights and biases are point estimates.}
    \label{fig:classical_nn}
    \end{subfigure}
    \hfill
\begin{subfigure}{0.47\textwidth}
    \centering
    \includegraphics[width=\textwidth]{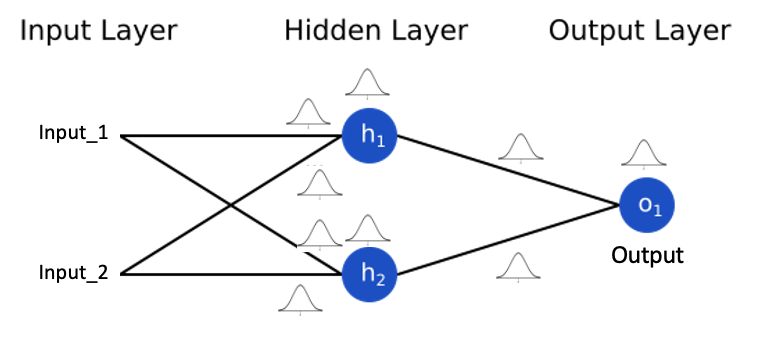}
    \caption{Bayesian Neural Network (BNN). Weights and biases are distributions.}
    \label{fig:BNN}
\end{subfigure}
\caption{Comparing a standard neural network to a BNN.}\label{fig:nn}
\end{figure}

Unlike classical deterministic neural networks, Bayesian Neural Networks (BNNs) are capable of making well-calibrated uncertainty predictions, which provide a measure of the uncertainty of the outcome \citep{jospin2020hands}. This is possible due to the fact that the weights and biases on at least some of the layers in the network are distributions rather than single point estimates (see Figure \ref{fig:nn}). More specifically, as BNNs use a Bayesian framework, once trained, the distributions of the weights and biases represent the posterior distributions based on the training data \citep{Bykov2020}. Note that for brevity in this section hereafter, we refer to the weights and biases as network parameters. The distributions in the output layer facilitate the assessment of aleatoric uncertainty (uncertainty in the data) and the distributions in the hidden layers facilitate the assessment of epistemic uncertainty (uncertainty in the model) \citep{salama_2021}. In this work, we choose to assess both types of uncertainty and use distributions for the output layer, as well as for the network parameters of the hidden layers. Our BNN approach therefore provides a more holistic view than previous work to assess uncertainty in large-scale ocean neural network predictions in \cite{gordon2022incorporating} where a deterministic neural network is used to predict the mean and variance of the output distribution.

Following \cite{jospin2020hands}, the posterior distributions in the BNN (\textit{i.e.} the distributions of the network parameters given the training data) are calculated using Bayes rule
\begin{equation}\label{eq:bates}
    p\left(W\lvert D_{tr}\right) = \frac{p(D_{tr}\lvert W)p(W)}{p({D_{tr})}} = \frac{p(D_{tr}\lvert W)p(W)}{\int_{W} p(D_{tr}\lvert W)p(W)\,dW},
\end{equation}
where $W$ are the network parameters, $D_{tr} = \left(x_{n}, y_{n}\right)$ the training data and $p(W)$ the prior distribution of the parameters. The probability of output $y$ given input $x$ is then given by the marginal probability distribution
\begin{equation}
    p(y\lvert x, D_{tr}) = \int_{W} p(y\rvert f(x; W))p(W\rvert D_{tr})\, dW,
\end{equation}
where $f(\cdot; W)$ is the neural network. However, computing $p\left(W\lvert D_{tr}\right)$ directly is very difficult, especially due to the denominator in (\ref{eq:bates}) which is intractable \citep{jospin2020hands, Bykov2020}. A number of methods have been proposed to approximate the denominator term including Markov Chain Monte Carlo sampling \citep{titterington2004bayesian} and variational inference \citep{osawa2019practical}. We use the latter which approximates the posterior using a variational distribution, $q_{\Phi}(W)$, with a known formula dependent on the parameters, $\Phi$, that define the distribution (for example for a normal distribution, $\Phi$ are its mean and variance). The BNN then learns the parameters $\Phi$ which lead to the closest match between the variational distribution and the posterior distribution \textit{i.e.} the parameters $\Phi$ which minimise the following Kullback–Leibler divergence (KL-divergence) 
\begin{equation}\label{KL-div}
    D_{KL}(q_{\Phi}\rvert\rvert p) = \int_{W} q_{\Phi}(W')\log\left(\frac{q_{\Phi}(W')}{p\left(W'\lvert D_{tr}\right)}\right)\,dW'.
\end{equation}
This formula still requires the posterior to be computed and so following standard practice, we use the ELBO formula instead
\begin{equation}\label{ELBO}
\int_{W} q_{\Phi}(W')\log\left(\frac{p(W', D_{tr})}{q_{\Phi}(W')}\right)\,dW',
\end{equation}
which is equal to $\log(p(D_{tr})) - D_{KL}(q_{\Phi}\rvert\rvert p)$. Thus maximising (\ref{ELBO}) is equivalent to minimising (\ref{KL-div}) since $\log(p(D_{tr}))$ only depends on the prior \citep{jospin2020hands}. In our work, we follow standard practice and assume that all variational forms of the posterior are normal distributions and thus the $\Phi$ parameters the neural network learns are the mean and variance of these distributions. Furthermore, for all priors in the BNN, we use the normal distribution $\mathcal{N}(0, 1)$, which is again standard practice because of the normal distribution's mathematical properties and simple log-form \citep{silvestro2020prior}. 

In our work, we also calculate the entropy of the final distribution as a measure of uncertainty. In information theory, entropy is considered as the expected information of a random variable and for each sample $i$ is given by
\begin{equation}\label{eq:entropy}
    H_{i} = - \sum_{j=1}^{N_{l}} p_{ij}\log(p_{ij}),
\end{equation}
where $N_{l}$ is the number of possible variable outcomes and $p_{ij}$ is the probability of each outcome $j$ for sample $i$ \citep{Goodfellow-et-al-2016}. Hence, the larger the entropy value, the less skewed the distribution and the more uncertain the model is of the result. 

Finally, for the layer architecture of the BNN, we use the same architecture as in \cite{THOR}, who use a deterministic neural network to predict ocean regimes from the same dataset as ours (see Section \ref{sec:data}). Thus, our BNN has 4 layers with [24, 24, 16, 16] nodes and `tanh' activation, where the layers are `DenseVariational' layers from the TensorFlow probability library \citep{dillon2017tensorflow}, rather than the `Dense' layers used in \cite{THOR}. For the output layer of the network, we use the `OneHotCategorical' layer from the TensorFlow probability library instead of a `SoftMax' layer and thus use the negative log-likelihood function as the loss function. The network is compiled with an Adam Optimizer \citep{kingma2014adam} with an initial learning rate of 0.01, which is reduced by a factor of 4 if the loss metric on the validation dataset does not decrease after 15 epochs (\textit{i.e.} after the entire training dataset has passed through the neural network fifteen times). The network is trained for 100 epochs and the best model network parameters over all epochs are recorded and saved as the trained parameters.

\subsection{Explainable AI (XAI)}\label{sec:XAI}
Whilst using a BNN enables scientists to determine how certain the network is of its results, being able to explain the source of the predictive skill is also of key importance particularly because of the potential for spurious correlations in neural networks giving rise to nonphysical predictions. As discussed in Section \ref{sec:intro}, XAI techniques have recently been developed to explain the skill of neural networks (\textit{i.e.} explain the correlations between the input features that give rise to the predictions). These techniques can then be used to reveal the extent to which neural networks are fit for purpose for a given problem \citep{samek2019explainable,arrieta2020explainable}. However, there has been little research into combining XAI techniques with uncertainty analysis. In this section, we outline how to adapt the two common XAI techniques, LRP and SHAP, so that they can be applied to BNNs. We remind the reader that we selected two XAI techniques originating from two different classes to gain a holistic view of the skill of the BNN. This is important to ensure that what the BNN has learned is genuinely rooted in physical theory, and we compare the outcomes of these methods with intuition from that theory.

\subsubsection{Layer-wise Relevance Propagation (LRP)}\label{LRP_method}
LRP explains network skill by calculating the contribution (or \emph{relevance}) of each input datapoint to the output score \citep{binder2016layer}. This leads to the construction of a `heatmap' where a positive/negative `relevance' means a feature contributes positively/negatively to the output \citep{bach2015pixel}. For a neural network, this relevance is calculated by back-propagating the relevance layer-by-layer from the output layer to the input layer. 

LRP has been successfully used to explain neural network skill in fields as diverse as medicine \citep{bohle2019layer}, information security \citep{seibold2020accurate} and text analysis \citep{arras2017relevant}, and has also already been applied to deterministic neural networks in climate science \citep{THOR,toms2020physically,Mamalakis2022}. However, there has been little research into applying LRP to BNNs, because the formulae used to calculate the relevance are difficult to apply when the network parameters are distributions. 

BNNs do however have the advantage that it is easy to generate a deterministic ensemble of networks from them, simply by sampling network parameters from the distributions. We therefore follow the novel methodology in \cite{Bykov2020} and use LRP on this ensemble of networks, efficiently generating an ensemble of LRP values which serve as a proxy for explaining the skill of the BNN. Each datapoint has its own distribution of LRP values and own level of uncertainty. If a datapoint has positive or negative relevance for every ensemble member, we can be increasingly confident about this point's relevance for explaining the skill of the BNN. For the remaining points, still following \citep{Bykov2020}, quantile heatmaps of the ensemble of LRP values can be used to visualise how many ensemble members have positive relevance and how many have negative.

There are many different formulae for calculating the relevance score with LRP \citep[see][]{montavon2019layer}, but in this work, we follow \cite{THOR} and use the LRP-$\epsilon$ rule which is good for handling noise. The relevance at layer $l$ of a neuron $i$ is then the sum of $R_{i \leftarrow j}^{(l, l+1)}$ for all neurons $j$ in layer $l+1$ where
\begin{equation}
R_{i \leftarrow j}^{(l, l+1)} = \frac{z_{ij}}{z_{j} + \epsilon \hspace{3pt}\text{sign}(z_{j})} R_{j}^{(l+1)}.
\end{equation}
Here $z_{ij}$ is the activation at neuron $i$ multiplied by the weight from neuron $i$ to $j$ and $z_{j} = \sum_{i}z_{ij}$ (see \cite{montavon2019layer} for more details).

\subsubsection{SHapley Additive exPlanation (SHAP) values}\label{sec:shap_method}
For our second XAI technique, we consider Shapley Additive Explanation values, known more commonly as SHAP values. These were first proposed in the context of game theory in \cite{shapley1997value}, but have since been extended to explaining skill in neural networks \citep{lundberg2017unified} and have been applied in climate science to deterministic neural networks in \cite{dikshit2021interpretable,Mamalakis2022}. There has been work adding uncertainty to the SHAP values of deterministic neural networks by adding noise \citep{slack2021reliable}, but this work represents the first time SHAP values are used to explain the skill of a BNN.

SHAP values are designed to compute the contribution of each input datapoint to the neural network output using a type of occlusion analysis. They test the effect of removing/adding a feature to the final output \textit{i.e.} calculating $f_{F}(x) - f_{F\backslash i}(x)$, where $f$ is the model, $F$ is the set of all features and $i$ the feature being considered \citep{lundberg2017unified}. To calculate the SHAP value, we must combine this for all features in the model with a weighted average meaning the SHAP value of feature $i$ for output $y = f_{F}(x)$ is
\begin{equation}\label{shap}
    \phi_{i}(x) = \sum_{S \subset F\backslash i} \frac{\lvert S\rvert ! (\lvert F\rvert - \lvert S\rvert -1)!}{\lvert F\rvert !}[f_{S \cup \{i\}}(x) - f_{S}(x)],
\end{equation}
where $S$ are all the sub-sets of $F$ excluding feature $i$. Note that summing the SHAP value for every feature $i$ gives the difference between the model prediction and the null model \textit{i.e.}
\begin{equation}
    f_{F}(x) = \mathbb{E}[y] + \sum_{i} \phi_{i}(x),
\end{equation}

where $\mathbb{E}[y]$ is the average of all outputs $y$ in the training dataset \citep{mazzanti2020shap}. We remark here that evaluating (\ref{shap}) for every feature can be computationally expensive; the complexity of the problem scales by $2^{\lvert F\rvert}$. Therefore various techniques have been proposed to speed up the evaluation of SHAP values, the most popular of which is KernelSHAP \citep{lundberg2017unified}. In this work, however, we choose to calculate the exact SHAP values because we only have eight features (see Section \ref{sec:data}) and these more efficient techniques assume feature independence (which our dataset does not have), and can lead to compromises on accuracy if not handled appropriately \citep{aas2021explaining}.

Like with LRP, we apply SHAP to an ensemble of deterministic neural networks generated from the BNN. We note here that SHAP is model agnostic so in the future, with changes to implementation, it may be possible to apply SHAP directly to the BNN itself. We expect the SHAP results to differ from the LRP results because the LRP ensemble captures the model uncertainty as LRP values are a weighted sum of the network weights, whereas SHAP captures the sensitivities of the outputs as a result of these uncertainties.

\section{Data}\label{sec:data}
\begin{figure}
    \centering
    \includegraphics[width=0.65\textwidth]{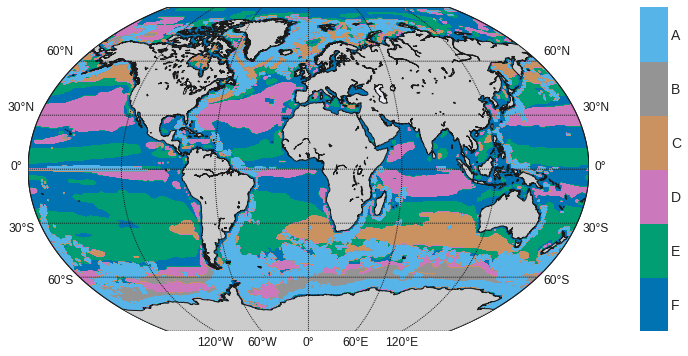}
    \caption{Global representation of dynamical ocean regimes in ECCOv4 data. For a full description of the ocean regimes see \cite{THOR}.}
    \label{fig:ecco_labels}
\end{figure}

A recent IPCC Special report highlights the need for a better understanding of uncertainty in ocean circulation patterns \citep{hoegh2018impacts}. An understanding of emergent circulation patterns can be gained using a dynamical regime framework \citep{sonnewald2019unsupervised}. These regimes simplify dynamics and each regime is then defined to be the solution space where the simplification is justifiable \citep{kaiser2021objective}. \cite{sonnewald2019unsupervised} show that unsupervised clustering techniques such as k-means clustering can be used to identify and partition dynamical regimes if the equations governing the dynamics are known. Specifically they use k-means clustering of model data from the numerical ocean model ECCOv4 (Estimating the Circulation and Climate of the Ocean) to identify dynamical regimes and develop geoscientific utility criteria. In our work, we follow \cite{THOR} and use this regime deconstruction framework as the labelled target data that the BNN seeks to predict at each point on the grid. Because the dynamical regimes were found in the model equation space, we have an automatic way to verify the explainable AI results. Figure \ref{fig:ecco_labels} shows a global representation of these six dynamical ocean regimes, which we have labelled A, B, C, D, E and F corresponding to the regimes `NL', `SO', `TR', `N-SV', `S-SV' and `MD' in \cite{THOR}. We have made this label simplification because the aim of this work is to develop a neural network technique to improve trustworthiness in ocean predictions. Thus anything other than a high-level understanding of the physics is beyond the scope of this work and we refer the reader to \cite{sonnewald2019unsupervised} and \cite{THOR} for a more in-depth discussion.

\begin{table}[H]
\scriptsize
\centering
\begin{tabular}{l|cccccc}
                            & \multicolumn{6}{c}{\textbf{Features}}   \\ 
\multirow{-2}{*}{\textbf{}} & \begin{tabular}[c]{@{}c@{}}Wind stress\\ curl\end{tabular} & Bathymetry                  & \begin{tabular}[c]{@{}c@{}}Dynamic\\ sea level\end{tabular} & Coriolis                    & \begin{tabular}[c]{@{}c@{}}Gradient \\ bathymetry\end{tabular} & \begin{tabular}[c]{@{}c@{}}Gradient \\ dynamic\\ sea level\end{tabular} \\ \hline
A  & {\color[HTML]{010066} High} & {\color[HTML]{010066} High} & {\color[HTML]{010066} High}   & {\color[HTML]{010066} High} & {\color[HTML]{010066} High}  & {\color[HTML]{010066} High}  \\
B & {\color[HTML]{010066} High}          & {\color[HTML]{010066} High} & {\color[HTML]{010066} High}                                 & {\color[HTML]{010066} High} & {\color[HTML]{010066} High}       & {\color[HTML]{010066} High} \\
C  & {\color[HTML]{010066} High}   & Med                         & Med   & {\color[HTML]{010066} High} & Med  & Med   \\
D   & {\color[HTML]{FE0000} Low}  & {\color[HTML]{FE0000} Low}  & {\color[HTML]{FE0000} Low}  & Med  & {\color[HTML]{FE0000} Low}   & {\color[HTML]{FE0000} Low}  \\
E   & Med    & Med   & Med    & {\color[HTML]{010066} High} & Med & Med  \\
F   & Med  & Med  & Med  & Med  & Med   & Med 
\end{tabular}
\caption{Approximate importance of features for predicting each regime according to the equation space, using analysis from Figure 1 in \cite{sonnewald2019unsupervised}.}\label{table:regimes}
\end{table}

 For our input features, we follow \cite{THOR} and use data from the numerical ocean model ECCOv4 (Estimating the Circulation and Climate of the Ocean), but the framework is set up so that it can be readily trained on CMIP6 data in the future \citep{forget2015ecco}. The following features are then used for prediction: wind stress curl, Coriolis (deflection effect caused by the Earth's rotation), bathymetry (measurement of ocean depth), dynamic sea level, and the latitudinal and longitudinal gradients of the bathymetry and the dynamic sea level. These features are chosen following the dynamical regime decomposition in \cite{sonnewald2019unsupervised} and Table \ref{table:regimes} shows which features are important for each regime according to the clustering of the equation space based on theoretical intuition. The specific composition of these features into terms in the equation space then manifests as different key ocean circulation patterns. Finally, for the training and test dataset split, we split by ocean basin and use shuffle for validation. The Atlantic Ocean basin ($80^{o}$W to $20^{o}$E) is the test dataset and the rest of the global ocean dataset is the training dataset.

\section{Results}\label{sec:results}
In this section, we first use a BNN to make a probabilistic forecast of ocean circulation regimes and show the value added by the uncertainty analysis that can be conducted through using a BNN instead of a deterministic neural network. We then use two modified XAI techniques to explain the skill of this network, comparing the two techniques with each other and with physical understanding.
\subsection{Bayesian Neural Networks (BNNs)}\label{subsec:results_bnn}
\begin{figure}
    \centering
    \includegraphics[width=0.47\textwidth]{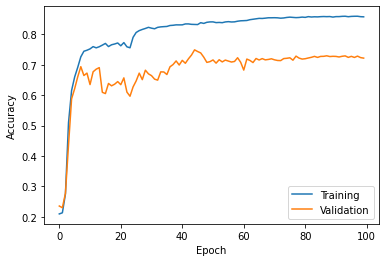}
\caption{Training accuracy and loss metrics for the BNN showing that the training has converged. Recall from Section \ref{sec:data} that the training dataset is the global ocean, excluding the Atlantic Ocean basin, and that shuffle is used for validation.}\label{fig:training}
\end{figure}

\begin{figure}
\begin{center}
\begin{subfigure}{0.4
\textwidth}
\centering
    \includegraphics[width=0.6\textwidth]{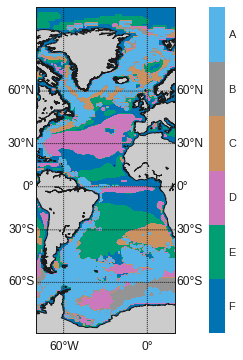}    \caption{Correct dynamical ocean regimes map.}
    \label{fig:regime_map}
    \end{subfigure}
\begin{subfigure}{0.4
\textwidth}
\centering
    \includegraphics[width=0.6\textwidth]{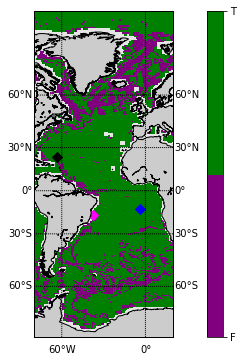}
    \caption{Accuracy (T = Correct; F = Incorrect).}
    \label{fig:correct_map}
    \end{subfigure}
    \end{center}
        \begin{subfigure}{0.32
\textwidth}
    \centering
    \includegraphics[height=\textwidth,width=\textwidth]{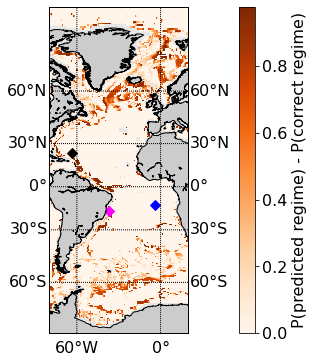}
    \caption{Difference between $\mathbb{P}$(predicted regime) and $\mathbb{P}$(correct regime).}
    \label{fig:probability_spatial}
    \end{subfigure}
    \hfill
\begin{subfigure}{0.32
\textwidth}
    \centering
    \includegraphics[height=\textwidth,width=\textwidth]{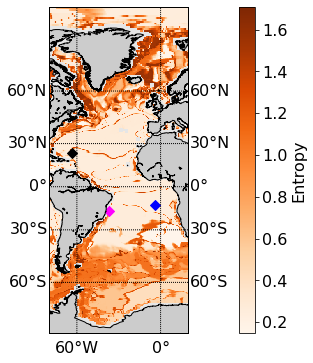}
    \caption{Entropy.\\\hspace{2pt}}
    \label{fig:entropy_spatial}
    \end{subfigure}
    \hfill
    \begin{subfigure}{0.32
\textwidth}
    \centering
    \includegraphics[height=\textwidth,width=\textwidth]{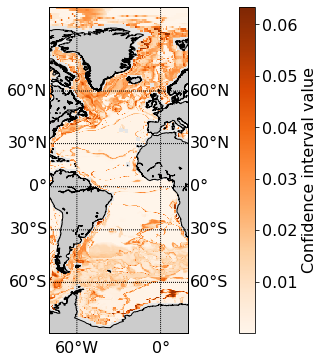}
    \caption{Confidence Interval value.\\\hspace{2pt}}
    \label{fig:ci_spatial}
    \end{subfigure}
    \caption{Spatial distribution of key metrics calculated from the BNN predictions for the test dataset (Atlantic Ocean basin), as well as the correct regimes in this region. The diamonds are the three locations of the example datapoints in Figure \ref{bin_examples}.}\label{fig:spatial}
\end{figure}

The advantage of BNNs over deterministic neural networks is the uncertainty estimate they provide. However, for BNNs to be of value they must also make accurate predictions. Figure \ref{fig:training} compares the accuracy metrics of the BNN applied to the training dataset (the global ocean, excluding the Atlantic Ocean basin) and the validation dataset (shuffled) during training. The accuracy metric clearly converges and the level of accuracy is high,  indicating that the architecture and learning rates chosen are appropriate for this dataset. When the trained BNN is applied to the test dataset (the Atlantic Ocean basin), the accuracy is 80\%, which is approximately the same as the accuracy achieved by the deterministic neural network in \cite{THOR} on the same data. Thus, by using a BNN we have not lost accuracy. Figure \ref{fig:correct_map} shows the spatial distribution of the correct and incorrect regime predictions. Most incorrect predictions occur for regime A for which errors are not unexpected -- it is a composite regime with a less Gaussian structure meaning it is less clearly defined and less easily determined by k-means \citep{sonnewald2019unsupervised}.

\begin{figure}
\centering 
\begin{subfigure}{0.47\textwidth}
    \centering
    \includegraphics[width=\textwidth]{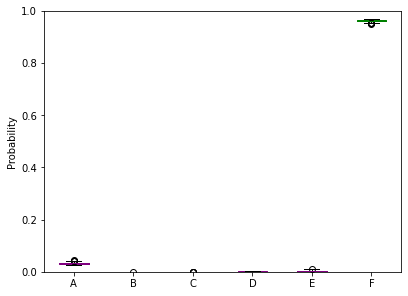}
    \caption{Example where correct regime predicted with high certainty (Location is blue diamond in Figure \ref{fig:spatial}).}
    \label{fig:corr_certain}
    \end{subfigure}
    \hfill
    \begin{subfigure}{0.47\textwidth}
    \centering
    \includegraphics[width=\textwidth]{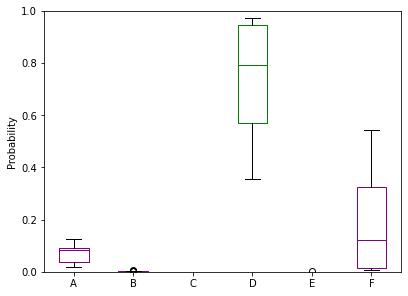}
    \caption{Example where correct regime predicted with some epistemic uncertainty (Location is black diamond in Figure \ref{fig:spatial}).}
    \label{fig:corr_less_certain}
    \end{subfigure}
    \begin{subfigure}{0.47\textwidth}
    \centering
    \includegraphics[width=\textwidth]{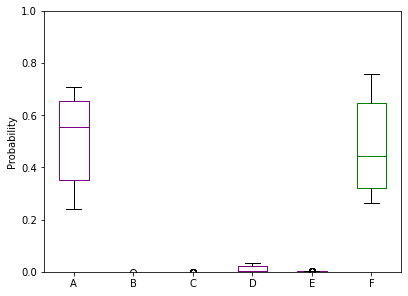}
    \caption{Example where incorrect regime predicted with both epistemic and aleatoric uncertainty (Location is magenta diamond in Figure \ref{fig:spatial}).}
    \label{fig:incorr}
    \end{subfigure}
    \caption{Box-and-whisker plot of BNN predictions of ocean regimes, generated using an ensemble of outputs. The correct regime is coloured green and the incorrect regimes are coloured purple.}\label{bin_examples}
\end{figure}

As we are considering aleatoric uncertainty (uncertainty in the input data), the BNN output is not deterministic but is instead a distribution. Moreover, as we are also considering epistemic uncertainty (uncertainty in the model parameters), the network parameters are distributions, the full output is an ensemble of distributions. In Figure \ref{bin_examples}, we show both types of uncertainty using a box-and-whisker plot for the predictions for three example datapoints. The narrower the box and whisker, the lower the epistemic uncertainty in the prediction for this regime. For example, in Figure \ref{fig:corr_certain} there is almost no  width to the box and whisker indicating low epistemic uncertainty, whereas for Figure \ref{fig:corr_less_certain} there are a range of possible probabilities of the most likely regime occurring, indicating epistemic uncertainty. In both Figures \ref{fig:corr_certain} and \ref{fig:corr_less_certain} the highest probability is high (almost 1 for Figure \ref{fig:corr_certain} and just under 0.8 on average for Figure \ref{fig:corr_less_certain}), which indicates that the aleatoric uncertainty is low. Therefore, practitioners can be confident in the results for both these datapoints, with Figure \ref{fig:corr_certain} being more trustworthy than Figure \ref{fig:corr_less_certain}. By contrast, Figure \ref{fig:incorr} has high levels of epistemic uncertainty and fairly high levels of aleatoric uncertainty meaning that although the practitioner can trust that the regime is either A or F, the overall regime prediction for this datapoint is not very trustworthy.

Using these distributions, we can calculate the difference between the probability the BNN assigns to the predicted regime and the probability it assigns to the correct regime. If the BNN has predicted the correct regime then this difference is zero, and, if the BNN is very certain in its prediction of the incorrect regime, the maximum possible probability difference is one. The spatial distribution of this value is shown in Figure \ref{fig:probability_spatial} and unsurprisingly corresponds closely with the spatial distribution of the correct and incorrect BNN predictions in Figure \ref{fig:correct_map}. The probability difference map adds value compared to the accuracy map because we can see where errors are more substantial. For example, although the BNN appears to perform poorly in the accuracy statistics around Greenland (especially around $50^{\circ}$W and $50^{\circ}$N and $20^{\circ}$W and $70^{\circ}$N), the difference between the probability of the correct regime and the highest probability is low. Therefore the BNN is still assigning a high probability to the correct regime here which is useful for practitioners. In contrast, off the north coast of South America, the probability difference is almost 1 meaning the BNN is doing a poor job here and should not be used in its current state for predictions here. Comparing Figure \ref{fig:probability_spatial} with Figure \ref{fig:regime_map} reveals that almost all the high probability differences occur at the boundaries between regime A and other regimes (for example in the Southern Ocean at the boundary between regimes B and D with regime A), indicating this is a weakness in the BNN. Thus by analysing this probability difference, we have gained valuable information for future predictions and learnt that to improve the BNN accuracy, we should provide more training data on the boundaries between regime A and other regimes.

\begin{figure}
    \centering
    \includegraphics[width=\textwidth]{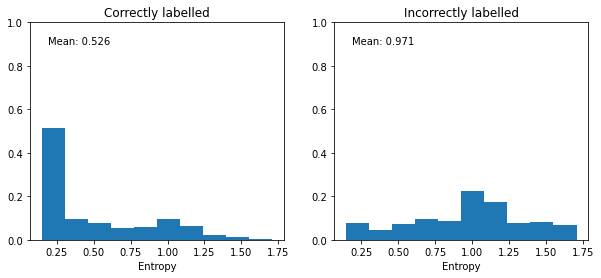}
    \caption{Distribution of entropy values for the correct and incorrect regime predictions. Recall that the lower the entropy, the more certain the result.}
    \label{fig:entropy}
\end{figure}

The distributions outputted by the BNN can also be used to numerically quantify the uncertainty in the network predictions. We can calculate the entropy value using (\ref{eq:entropy}), where we recall that the higher the value the more uncertain the result. Figure \ref{fig:entropy_spatial} shows the spatial distribution of this entropy and comparing with Figure \ref{fig:correct_map} shows that the higher entropy values tend to be where the BNN prediction is incorrect. More precisely, Figure \ref{fig:entropy} compares the distribution of the entropy when the BNN predictions are correct and when they are incorrect, and clearly shows that the entropy for the correct predictions is skewed towards lower values, whereas the entropy for the incorrect predictions is skewed higher. This is a good result because it means that the predictions are notably more uncertain when they are incorrect than when they are correct, \textit{i.e.} the correct results are also the results that the BNN informs the practitioner are the most trustworthy.

\begin{figure}
\begin{subfigure}{0.4\textwidth}
    \centering
    \includegraphics[width=0.6\textwidth]{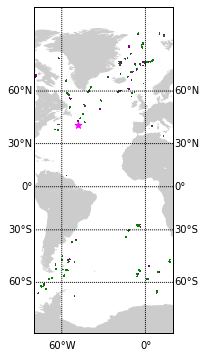}
    \caption{Spatial distribution of points where the most likely regime is not statistically significantly different from other regimes.}
    \label{fig:stat_sign}
\end{subfigure}
\hfill
\begin{subfigure}{0.4\textwidth}
    \centering
    \includegraphics[width=0.8\textwidth]{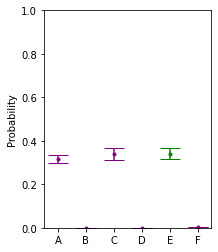}
    \caption{Confidence interval plot of example datapoint, where the probabilities for the top three regimes are not statistically significantly different.}
    \label{fig:cont_int_eg}
\end{subfigure}
\caption{Considering whether the differences between the probabilities for each regime are statistically significantly different. The star on (a) is the location of the example datapoint in (b). In both figures, incorrect predictions are coloured purple and correct predictions green.}
\end{figure}

Finally, Figure \ref{bin_examples} show that there can be substantial overlap between the box-and-whisker for each regime. However this can be misleading as box-and-whisker plots consider upper and lower quartiles which are not useful for assessing statistical significance. Therefore, we also consider the confidence intervals and in Figure \ref{fig:ci_spatial} show the spatial distribution of their size. Note that unsurprisingly, the spatial distribution for the confidence intervals is very similar to that for the entropy because they are calculated using similar statistics. Using confidence intervals, we find that for the majority of cases, the probabilities for the most likely regime are statistically significantly different from the probabilities for the other regimes. Figure \ref{fig:stat_sign} highlights the datapoints for which this is not the case, and unsurprisingly shows these datapoints correspond to points for which there is high entropy (see Figure \ref{fig:entropy_spatial}). For the vast majority of the datapoints in Figure \ref{fig:stat_sign}, the top two most likely regimes are statistically significantly different from the other regimes and the correct regime is one of the two regimes. Therefore although the neural network is uncertain for these datapoints, it is still predicting a high probability for the correct regime. Finally, there are approximately 20 datapoints where only the top three most likely regimes are significantly different from the others. An example of one such datapoint is shown in Figure \ref{fig:cont_int_eg}, where half the regimes have the same probability. Although this is not ideal, this is an example of where a BNN is better than a deterministic neural network, because it clearly informs the user that it is very uncertain of its prediction and that using this BNN on this datapoint is inappropriate.

Therefore, in this section we have shown that by looking at the probabilities and confidence intervals produced by the BNN, practitioners can make an informed decision as to whether to trust the BNN prediction for the dynamical regime or whether further analysis is required for these datapoints.

\subsection{Explainable AI (XAI)} 
To explain the BNN's skill, we apply two common XAI techniques, LRP and SHAP, to an ensemble of deterministic neural networks generated from the BNN. We consider LRP in Section \ref{LRP_results} and SHAP in Section \ref{sec:shap_results}, and then compare results from the two techniques in Section \ref{sec:shap_vs_LRP} to test the `disagreement problem' discussed in Section \ref{sec:XAI}. If LRP and SHAP largely agree with each other as to which features are relevant in each region (\textit{i.e.} there is no disagreement problem) and also agree with our intuition from physical theory then this increases the trust in our XAI results. This is important to ensure that what the BNN has learned is genuinely rooted in physics.' Moreover, the use of a BNN allows us to explore whether disagreement between SHAP and LRP is more likely to occur when predictions have higher entropy (\textit{i.e.} higher uncertainty).
\subsubsection{Layer-wise Relevance Propagation (LRP)}\label{LRP_results}
\begin{figure}
\begin{minipage}{\textwidth}
\centering
    \includegraphics[width=0.25\textwidth]{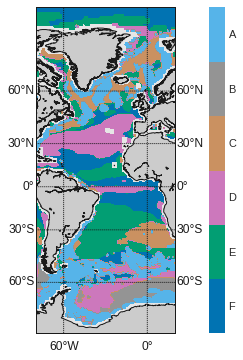}   
    \caption{Most probable ocean regime predicted by Bayesian Neural Network.}\label{fig:predicted_regimes}
    \end{minipage}
\begin{minipage}{\textwidth}
\begin{subfigure}{0.24\textwidth}
    \centering
    \includegraphics[height=1.5\textwidth]{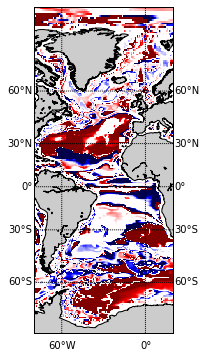}
    \caption{Wind stress curl.}
    \label{fig:wind_stress_LRP}
    \end{subfigure}
            \hfill
        \begin{subfigure}{0.24\textwidth}
    \centering
    \includegraphics[height=1.5\textwidth]{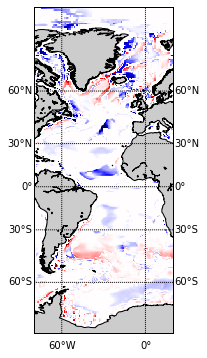}
    \caption{Bathymetry.}
    \label{fig:bath_LRP}
    \end{subfigure}
    \hfill
\begin{subfigure}{0.24\textwidth}
    \centering
    \includegraphics[height=1.5\textwidth]{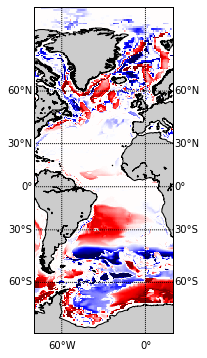}
    \caption{Dynamic sea level.}
    \label{fig:ssh_LRP}
    \end{subfigure}
    \hfill
  \begin{subfigure}{0.24\textwidth}
    \centering
    \includegraphics[height=1.5\textwidth]{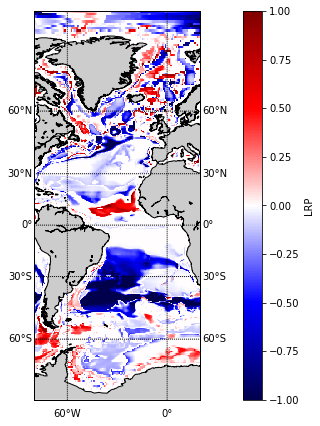}
    \caption{Coriolis force.}
    \label{fig:ssh_F}
    \end{subfigure}  
    
    \begin{subfigure}{0.24\textwidth}
    \centering
    \includegraphics[height=1.5\textwidth]{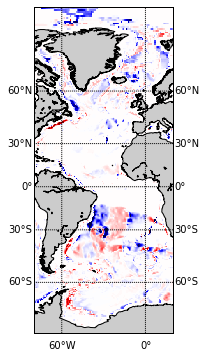}
    \caption{Gradient bathymetry (lon).}
    \label{fig:bathx_LRP}
    \end{subfigure}
        \hfill
\begin{subfigure}{0.24\textwidth}
    \centering
    \includegraphics[height=1.5\textwidth]{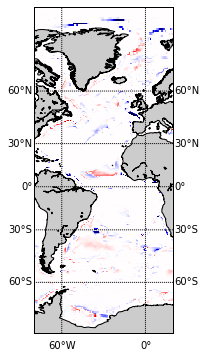}
        \caption{Gradient bathymetry (lat).}
    \label{fig:bathy_LRP}
\end{subfigure}
\hfill
\begin{subfigure}{0.24\textwidth}
    \centering
    \includegraphics[height=1.5\textwidth]{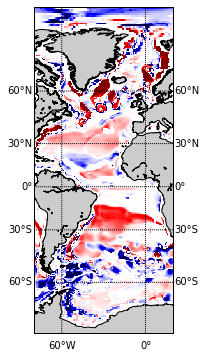}
    \caption{Gradient dynamic sea level (lon).}
    \label{fig:sshx_LRP}
    \end{subfigure}
        \hfill
\begin{subfigure}{0.24\textwidth}
    \centering
    \includegraphics[height=1.5\textwidth]{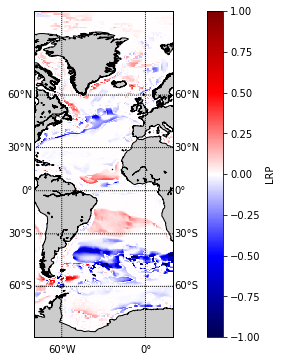}
        \caption{Gradient dynamic sea level (lat).}
    \label{fig:sshy_LRP}
    \end{subfigure}
    \caption{LRP values which are consistent across the whole ensemble. Red indicates that the variable in this area is actively helpful, blue that it is actively unhelpful, and white that it is too uncertain to have consistent relevance.}\label{LRP_consistent}
    \end{minipage}
\end{figure}

Applying LRP using our ensemble approach means that each input variable has its own distribution of LRP values and own level of uncertainty. Figure \ref{LRP_consistent} shows the values for which the sign of the LRP value (\textit{i.e} the relevance) remains the same between the 25\% to 75\% quantiles of the ensemble. Note that throughout the LRP values are scaled by the maximum absolute LRP value for any variable across the ensemble. If the LRP value consistently has the same sign across the quantiles, then we can be confident of the effect this feature has on the output; the piece of information of most interest to practitioners in a recent survey in \cite{lakkaraju2022rethinking}.

\begin{table}
\tiny
\begin{tabular}{l|cccccccccccccc}
                            & \multicolumn{14}{c}{\scriptsize{\textbf{Features}}} \\ 
\multirow{-2}{*}{\textbf{}} & \multicolumn{2}{c|}{\scriptsize{\begin{tabular}[c]{@{}c@{}}Wind stress\\ curl\end{tabular}}}    & \multicolumn{2}{c|}{ \scriptsize{Bathymetry}}                                              & \multicolumn{2}{c|}{\scriptsize{\begin{tabular}[c]{@{}c@{}}Dynamic\\ sea level\end{tabular}}}   & \multicolumn{2}{c|}{\scriptsize{Coriolis}}                            & \multicolumn{2}{c|}{\scriptsize{\begin{tabular}[c]{@{}c@{}}Gradient \\ bathymetry\end{tabular}}} & \multicolumn{2}{c|}{\scriptsize{\begin{tabular}[c]{@{}c@{}}Gradient \\ dynamic\\ sea level\\ (lon)\end{tabular}}} & \multicolumn{2}{c}{\scriptsize{\begin{tabular}[c]{@{}c@{}}Gradient \\ dynamic\\ sea level\\ (lat)\end{tabular}}}                                              \\ \cline{2-15} 
                             & Var                         & \multicolumn{1}{c|}{Rel.}                             & Var                        & \multicolumn{1}{c|}{Rel.}                       & Var                        & \multicolumn{1}{c|}{Rel.}                             & Var                         & \multicolumn{1}{c|}{Rel.}                             & Var                            & \multicolumn{1}{c|}{Rel.}                          & Var                                  & \multicolumn{1}{c|}{Rel.}                                     & Var                         & Rel.                                                                                            \\ \hline
A                           & {\color[HTML]{333333} Med}  & \multicolumn{1}{c|}{Med --} & Med & \multicolumn{1}{c|}{Med +} & {\color[HTML]{333333} Med} & \multicolumn{1}{c|}{{\color[HTML]{010066} High +}} & {Med}  & \multicolumn{1}{c|}{{Med +}}  & {\color[HTML]{FE0000} Low}     & \multicolumn{1}{c|}{{\color[HTML]{009901} Low}}    & {Med}           & \multicolumn{1}{c|}{{\color[HTML]{FE0000} High --}}         & {Med}  & {Med --}                                                                 \\
B                           & {\color[HTML]{FE0000} Low}  & \multicolumn{1}{c|}{{\color[HTML]{010066} High +}} & {\color[HTML]{FE0000} Low} & \multicolumn{1}{c|}{{Med --}} & {Med} & \multicolumn{1}{c|}{{\color[HTML]{010066} High +}} & {\color[HTML]{FE0000} Low}     & \multicolumn{1}{c|}{{\color[HTML]{009901} Low}} & {\color[HTML]{FE0000} Low}     & \multicolumn{1}{c|}{{\color[HTML]{009901} Low}}    & {\color[HTML]{FE0000} Low}           & \multicolumn{1}{c|}{{\color[HTML]{009901} Low}}               & {\color[HTML]{FE0000} Low}  & {\color[HTML]{009901} Low}                                                                      \\
C                           &  Med  & \multicolumn{1}{c|}{{\color[HTML]{010066} High +}} & {\color[HTML]{FE0000} Low} & \multicolumn{1}{c|}{{\begin{tabular}[c]{@{}c@{}}Med -- (NH)\\ Med + (SH)\end{tabular}}} & {\color[HTML]{FE0000} Low} & \multicolumn{1}{c|}{{\color[HTML]{009901} Low}}       & {\color[HTML]{FE0000} Low}  & \multicolumn{1}{c|}{{\color[HTML]{009901} Low}}       & {\color[HTML]{FE0000} Low}     & \multicolumn{1}{c|}{{\color[HTML]{009901} Low}}    & {\color[HTML]{FE0000} Low}           & \multicolumn{1}{c|}{{\color[HTML]{009901} Low}}               & {\color[HTML]{FE0000} Low}  & {\color[HTML]{009901} Low}                                                                      \\
D                           & Med  & \multicolumn{1}{c|}{{\color[HTML]{010066} High +}} & {Med} & \multicolumn{1}{c|}{{Med --}} & {\color[HTML]{FE0000} Low} & \multicolumn{1}{c|}{{\begin{tabular}[c]{@{}c@{}}Med + (NH)\\ Med -- (SH)\end{tabular}}}       & {Med}  & \multicolumn{1}{c|}{{Med --}}  & {\color[HTML]{FE0000} Low}     & \multicolumn{1}{c|}{{\color[HTML]{009901} Low}}    & {Med}           & \multicolumn{1}{c|}{{Med +}}          & {\color[HTML]{FE0000} Low}  & {\color[HTML]{009901} Low}                                                                      \\
E                           & {\color[HTML]{010066} High} & \multicolumn{1}{c|}{{\color[HTML]{010066} High +}} & {\color[HTML]{FE0000} Low} & \multicolumn{1}{c|}{{\color[HTML]{009901} Low}} & {\color[HTML]{FE0000} Low} & \multicolumn{1}{c|}{{\color[HTML]{010066} High +}} & {Med}  & \multicolumn{1}{c|}{{\color[HTML]{FE0000} High --}} & {\color[HTML]{FE0000} Low}     & \multicolumn{1}{c|}{{\color[HTML]{009901} Low}}    & {\color[HTML]{FE0000} Low}           & \multicolumn{1}{c|}{{\color[HTML]{010066} High +}}         & {\color[HTML]{FE0000} Low}  & {Med +}                                                                 \\
F                           & Med  & \multicolumn{1}{c|}{{Med --}}  & {Med} & \multicolumn{1}{c|}{{Med --}} & {\color[HTML]{FE0000} Low} & \multicolumn{1}{c|}{Med --}       & {\color[HTML]{010066} High} & \multicolumn{1}{c|}{{Med --}}  & {\color[HTML]{FE0000} Low}     & \multicolumn{1}{c|}{{\color[HTML]{009901} Low}}    & {\color[HTML]{010066} High}          & \multicolumn{1}{c|} {Med +}        & {\color[HTML]{010066} High} & { \begin{tabular}[c]{@{}c@{}}Med \\ (-- \textgreater +)\end{tabular}}
\end{tabular}
\caption{General trends in the variance and relevance of LRP values for each regime and each feature. Here + indicates that the feature is actively helpful and -- that it is actively unhelpful (so High + indicates high positive relevance). Note (-- \textgreater +) indicates that between the 25th and 75th quantiles, the variable changes from unhelpful to helpful.}\label{table:LRP_summary}
\end{table}

In Figure \ref{LRP_consistent}, red indicates that the variable in this area is actively helpful for the BNN in making its predictions, blue that it is actively unhelpful, and white that it is too uncertain to have consistent relevance. Note that certain areas of white may also be because the variable does not contribute (see Figure \ref{fig:LRP_quantile} in \ref{appendix} which shows the actual LRP values for the 25\%, 50\% and 75\% quantiles of the ensemble). An important point to note when interpreting these trends is that our network predicts using a gridpoint-by-gridpoint approach and does not see the overall global map, thus making the spatial coherence striking in its consistency. To aid with the interpretation of the LRP values for each regime, we include Figure \ref{fig:predicted_regimes} (which shows the most probable ocean regime predicted by the BNN) to help qualitatively see the trends, and Table \ref{table:LRP_summary} which highlights the general trends in the relevance and variance of the LRP values for each regime with respect to each feature. By comparing Table \ref{table:regimes} with Table \ref{table:LRP_summary}, we can compare the general trends of the LRP values with what is expected from the clustering of the equation space. A strong difference is that according to LRP the gradients of the bathymetry are irrelevant to the BNN predictions with high certainty (apart from in key regions discussed in Table \ref{table:LRP_features}), whereas the equation space suggests the bathymetry gradients are relevant for some regimes.

Of particular interest when comparing Tables \ref{table:regimes} with \ref{table:LRP_summary} are the differences for Regimes A and B. From the equation space (see Table \ref{table:regimes}), we would expect all features to be actively helpful for these regimes. However, in the case of Regime A, the LRP values conclude that both the wind stress curl and the longitudinal gradient of the dynamic sea level are actively unhelpful. Figure \ref{fig:spatial} shows that both the highest areas of inaccuracy and the highest areas of entropy (\textit{i.e} uncertainty) in the BNN occur for Regime A. These LRP values suggest that the reason for these errors and uncertainty is that the BNN is not correctly weighting the wind stress curl and the longitudinal gradient of the dynamic sea level for Regime A. By contrast, for Regime B, there are no features which are actively unhelpful. Instead, there are some features for which the BNN has no relevance (gradients of both the bathymetry and the dynamic sea level). The BNN predictions for Regime B are generally accurate and certain, and therefore this implies that, despite the conclusions from the equation space, the BNN can rely on certain key features it has identified to make accurate certain predictions. There is therefore scope for learning about the physical ocean processes guided by understanding of what the BNN determines as important and unimportant.

\begin{figure}
    \centering
    \includegraphics[width = 0.45\textwidth]{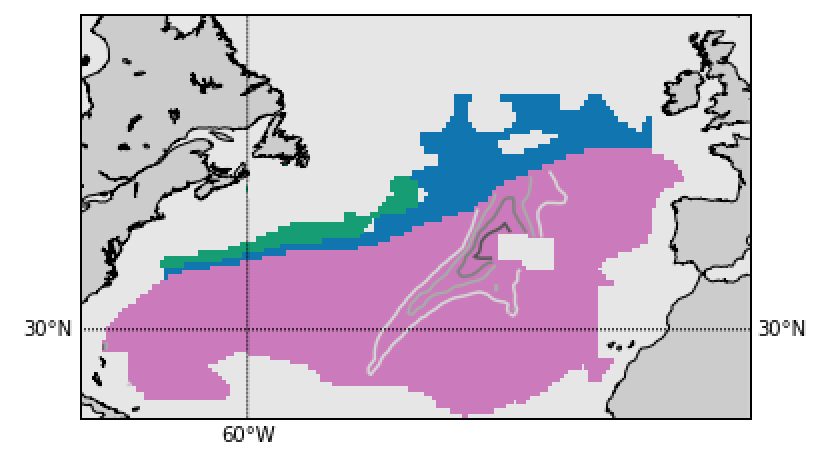}
    \caption{Locations of key dynamical processes and physical features of interest in Table \ref{table:LRP_features}: the North Atlantic Drift is the blue region at $\sim 40^{\circ}$N; the Gulf Stream leaving the continental shelf is the green region near coastline at $\sim 70^{\circ}$W and $40^{\circ}$N; the wind gyre is the pink region at $\sim 0^{\circ}$ and $30^{\circ}$S; and the part of the Mid-Atlantic Ridge we are focusing on is are the gray-scale contours crossing the wind gyre at $\sim 30^{\circ}$W.}  
    \label{fig:feature_map}
\end{figure}

\begin{table}
\scriptsize
\begin{tabular}{l|cccccccccccccc}
                                                                 & \multicolumn{14}{c}{\textbf{Features}}   \\ 
\multicolumn{1}{c|}{}                                            & \multicolumn{2}{c|}{\begin{tabular}[c]{@{}c@{}}Wind stress\\ curl\end{tabular}}                                      & \multicolumn{2}{c|}{Bath.}                                              & \multicolumn{2}{c|}{\begin{tabular}[c]{@{}c@{}}Dynamic\\ Sea Level\end{tabular}} & \multicolumn{2}{c|}{Coriolis}                                                    & \multicolumn{2}{c|}{\begin{tabular}[c]{@{}c@{}}Gradient\\ bath.\end{tabular}} & \multicolumn{2}{c|}{\begin{tabular}[c]{@{}c@{}}Gradient\\ sea level\\ (lon)\end{tabular}} & \multicolumn{2}{c}{\begin{tabular}[c]{@{}c@{}}Gradient\\ sea level\\ (lat)\end{tabular}} \\ \cline{2-15} 
\multicolumn{1}{c|}{}                                            & Var                         & \multicolumn{1}{c|}{Rel.}                                                              & Var                        & \multicolumn{1}{c|}{Rel.}                       & Var                         & \multicolumn{1}{c|}{Rel.}                          & Var                        & \multicolumn{1}{c|}{Rel.}                           & Var                          & \multicolumn{1}{c|}{Rel.}                           & Var                                  & \multicolumn{1}{c|}{Rel.}                                    & Var                                            & Rel.                                              \\ \hline
\begin{tabular}[c]{@{}l@{}}NAD\end{tabular} & {\color[HTML]{FE0000} Low}  & \multicolumn{1}{c|}{Med +}                                     & {\color[HTML]{FE0000} Low} & \multicolumn{1}{c|}{{\color[HTML]{009901} Low}} & Med                         & \multicolumn{1}{c|}{Med +}                         & {\color[HTML]{FE0000} Low} & \multicolumn{1}{c|}{{\color[HTML]{FE0000} High --}} & {\color[HTML]{FE0000} Low}   & \multicolumn{1}{c|}{{\color[HTML]{009901} Low}}     & Med          & \multicolumn{1}{c|}{Med --}              & {\color[HTML]{FE0000} Low}                     & {\color[HTML]{FE0000} High --}                    \\
\begin{tabular}[c]{@{}l@{}}GS\end{tabular}      & Med  & \multicolumn{1}{c|}{{\color[HTML]{010066} High +}}                                       & Med                        & \multicolumn{1}{c|}{Med --}                      & {\color[HTML]{FE0000} Low}  & \multicolumn{1}{c|}{{\color[HTML]{009901} Low}}   & Med & \multicolumn{1}{c|}{{\color[HTML]{010066} High +}} & Med   & \multicolumn{1}{c|}{Med --}  & {\color[HTML]{010066} High}         & \multicolumn{1}{c|}{{\begin{tabular}[c]{@{}c@{}}High\\ ({\color[HTML]{FE0000} --} \textgreater {\color[HTML]{010066} +})\end{tabular}}}           & Med                     & Med +                    \\
Gyre                         & {\color[HTML]{FE0000} Low} & \multicolumn{1}{c|}{{\color[HTML]{010066} High +}} & {Med} & \multicolumn{1}{c|}{{Med --}}                                                           & {\color[HTML]{FE0000} Low}  & \multicolumn{1}{c|}{{\color[HTML]{009901} Low}}    & {Med} & \multicolumn{1}{c|}{{\color[HTML]{FE0000} High --}} & {\color[HTML]{FE0000} Low}     & \multicolumn{1}{c|}{{\color[HTML]{009901} Low}}    & {Med}            & \multicolumn{1}{c|}{Med +}           & {\color[HTML]{FE0000} Low}                       & {\color[HTML]{009901} Low} 
                       \\ \hline 
\begin{tabular}[c]{@{}l@{}}MAR\end{tabular}   & {\color[HTML]{010066} High} & \multicolumn{1}{c|}{\begin{tabular}[c]{@{}c@{}}Med\\ (-- \textgreater +)\end{tabular}} &     {\color[HTML]{FE0000} Low}          & \multicolumn{1}{c|}{{\color[HTML]{FE0000} High --}}       & Med                         & \multicolumn{1}{c|}{Med --}                        & Med                        & \multicolumn{1}{c|}{{\color[HTML]{FE0000} High --}}         & Med                          & \multicolumn{1}{c|}{Med --}                         & Med          & \multicolumn{1}{c|}{{\color[HTML]{010066} High +}}           & {\color[HTML]{010066} High}                          & Med +                                            
\end{tabular}
\caption{Variance and relevance of LRP values for the key dynamical processes of the North Atlantic Drift (NAD); the Gulf Stream leaving the continental shelf (GS), the wind gyre and the key physical feature of the Mid-Atlantic Ridge as it crosses the wind gyre (MAR) (see Figure \ref{fig:feature_map}). Here + indicates that the feature is actively helpful and -- that it is actively unhelpful (so High + indicates high positive relevance). Note (-- \textgreater +) indicates that between the 25th and 75th quantiles, the variable changes from unhelpful to helpful.}\label{table:LRP_features}
\end{table}

For reasons of brevity, we do not detail all the physical interpretations in Figure \ref{LRP_consistent} and Table \ref{table:LRP_summary} but instead focus on the key dynamical processes of the North Atlantic Drift, the Gulf Stream leaving the continental shelf, and the North Atlantic wind gyre; and the key physical characteristic of the mid-Atlantic ridge specifically as it crosses the wind gyre (hereafter simply referred to as the mid-Atlantic ridge). The location of these processes is shown in Figure \ref{fig:feature_map} and the variance and relevance of the LRP values in these regions are summarised in Table \ref{table:LRP_features}. The table highlights that for the North Atlantic Drift, there are no features which have strong positive relevance; in fact, the Coriolis force and latitudinal gradient of the sea level have strong negative relevance.  Instead, the highly relevant areas for this region are not for the regime of the North Atlantic Drift (Regime F), but for the other regimes, for example, both the dynamic sea level and its longitudinal gradient are strongly positively relevant for Regime A in this region. This is also noted in \cite{THOR}, who suggest this could be because of multiple inputs contributing medium importance to predictions for Regime F (see Table \ref{table:regimes}). In contrast, where the Gulf Stream leaves the continental shelf, the Coriolis effect and wind stress curl are both strongly helpful. This conclusion greatly agrees with physical intuition, which states that these features are the key drivers for the Gulf Stream's movement across the North Atlantic \citep{webb2021introduction}. Table \ref{table:LRP_features} also shows that the bathymetry gradient is unhelpful for this process. Before leaving the coast, physical intuition suggests that the gradient of the bathymetry is the key driver of the Gulf Stream and this can be seen in the LRP values, (particularly for the latitudinal gradient in Figure \ref{fig:LRP_quantile}h). It is therefore likely that the BNN is using the same weightings for the bathymetry gradient as the Gulf Stream leaves the continental shelf, but the key drivers have changed meaning the bathymetry gradient is no longer helpful. Also of interest is the longitudinal gradient of the sea level, which is unhelpful for the North Atlantic Drift, very uncertain for the Gulf Stream leaving the continental shelf (a region which has high entropy in Figure \ref{fig:entropy_spatial}) and then helpful for the wind gyre. This suggests the this feature is acting as an indicator between the three regimes discussed here. For the wind gyre, the wind stress curl is also strongly helpful, which agrees with the intuition from physical theory of gyres, which states that they are largely driven by the wind stress curl \citep[see][]{munk1950wind}. Note however that the theory also indicates that Coriolis should be somewhat helpful but it is actively unhelpful. This variation may be because the BNN does not seem to be able to accurately weight low values of Coriolis (near the equator). Nevertheless the general agreement with physical intuition for the dynamical processes discussed here highlights our BNN's ability to learn key physical processes. 

Unlike the other processes highlighted, the mid-Atlantic ridge is a physical characteristic of the bathymetry that will remain unchanged by a future climate. The ridge is clearly identifiable in the features in Figure \ref{LRP_consistent} and it is therefore interesting to highlight the differences between the relevance of this ridge and the relevance of the other gridpoints in the wind gyre around it. The most noticeable difference is that the ridge adds uncertainty to the BNN predictions -- for almost all features, the relevance of the mid-Atlantic ridge is more uncertain than that of the wind gyre. The exception is the bathymetry, which becomes strongly unhelpful with high certainty at the mid-Atlantic ridge. Added to the fact that the bathymetry gradients are also more unhelpful at the ridge than at the surrounding gridpoints, this suggests that the BNN is able to identify the ridge in the bathymetry but unable to weight it correctly, which leads to uncertainty in the relevance of the other features. We observe that, in contrast to bathymetry, both gradients of the dynamic sea level increase in helpfulness at the ridge, in particular the longitudinal gradient. Moreover, Figure \ref{fig:spatial} shows the BNN predicts the correct regime for the mid-Atlantic ridge with high certainty. Therefore, this suggests that reliable and accurate predictions for regimes at the mid-Atlantic ridge should be based more on the gradient of the dynamic sea level than the bathymetry itself.

To summarise, our discussion of LRP values in this section has highlighted both our BNN's ability to identify known physical characteristics and the potential scope to advance physical theory through analysing its skill.

\subsubsection{SHapley Additive exPlanation (SHAP) Values}\label{sec:shap_results}
Whereas LRP considers the relevance of a feature for all regimes simultaneously, the SHAP approach sees the problem as binary for each regime: including a feature at a gridpoint either increases the probability of the specific regime being considered there or decreases it. There is therefore a SHAP value for each gridpoint for each regime, meaning we have six times the number of SHAP values as we do LRP. Moreover our ensemble approach means each input variable and regime pairing has its own distribution of SHAP values and own level of uncertainty. Table \ref{table:shap_summary} summarises the general trends in the SHAP values and in particular highlights that for all regimes and features the variance in the ensemble is low, and most features considered are actively helpful. The main exceptions to the latter are the latitudinal gradient of the dynamic sea level and both bathymetry gradients, which are not important for regime predictions (apart from in certain key areas discussed later).

\begin{table}[H]
\scriptsize
\begin{tabular}{l|cccccccccccccc}
                            & \multicolumn{14}{c}{\textbf{Features}}                                       \\
\multirow{-2}{*}{\textbf{}} & \multicolumn{2}{c|}{\begin{tabular}[c]{@{}c@{}}Wind stress\\ curl\end{tabular}} & \multicolumn{2}{c|}{Bathymetry}                                                                                                           & \multicolumn{2}{c|}{\begin{tabular}[c]{@{}c@{}}Dynamic\\ sea level\end{tabular}}                                                          & \multicolumn{2}{c|}{Coriolis}                                                   & \multicolumn{2}{c|}{\begin{tabular}[c]{@{}c@{}}Gradient \\ bathymetry\end{tabular}} & \multicolumn{2}{c|}{\begin{tabular}[c]{@{}c@{}}Gradient \\ sea level\\ (lon)\end{tabular}} & \multicolumn{2}{c}{\begin{tabular}[c]{@{}c@{}}Gradient \\ sea level\\ (lat)\end{tabular}} \\ \cline{2-15} 
                            & Var                        & \multicolumn{1}{c|}{Rel.}                          & Var                        & \multicolumn{1}{c|}{Rel.}                                                                                    & Var                        & \multicolumn{1}{c|}{Rel.}                                                                                    & Var                        & \multicolumn{1}{c|}{Rel.}                          & Var                            & \multicolumn{1}{c|}{Rel.}                          & Var                                   & \multicolumn{1}{c|}{Rel.}                                    & Var                                              & Rel.                                             \\ \hline
A                           & {\color[HTML]{FE0000} Low} & \multicolumn{1}{c|}{{Med +}}  & {\color[HTML]{FE0000} Low} & \multicolumn{1}{c|}{{\color[HTML]{010066} High +}}                                                           & {\color[HTML]{FE0000} Low} & \multicolumn{1}{c|}{{\color[HTML]{010066} High +}}                                                           & {\color[HTML]{FE0000} Low} & \multicolumn{1}{c|}{{Med +}}  & {\color[HTML]{FE0000} Low}     & \multicolumn{1}{c|}{{\color[HTML]{009901} Low}}    & {\color[HTML]{FE0000} Low}            & \multicolumn{1}{c|}{{\color[HTML]{010066} High +}}           & {\color[HTML]{FE0000} Low}                       & {\color[HTML]{009901} Low}                       \\
B                           & {\color[HTML]{FE0000} Low} & \multicolumn{1}{c|}{{\color[HTML]{010066} High +}} & {\color[HTML]{FE0000} Low} & \multicolumn{1}{c|}{{Med +}}                                                            & {\color[HTML]{FE0000} Low} & \multicolumn{1}{c|}{{\color[HTML]{010066} High +}}                                                           & {\color[HTML]{FE0000} Low} & \multicolumn{1}{c|}{{\color[HTML]{010066} High +}}   & {\color[HTML]{FE0000} Low}     & \multicolumn{1}{c|}{{\color[HTML]{009901} Low}}    & {\color[HTML]{FE0000} Low}            & \multicolumn{1}{c|}{{\color[HTML]{009901} Low}}            & {\color[HTML]{FE0000} Low}                       & {\color[HTML]{009901} Low}                       \\
C                           & {\color[HTML]{FE0000} Low} & \multicolumn{1}{c|}{{\color[HTML]{010066} High +}} & {\color[HTML]{FE0000} Low} & \multicolumn{1}{c|}{{\begin{tabular}[c]{@{}c@{}}Med -- (NH)\\ Med + (SH)\end{tabular}}} & {\color[HTML]{FE0000} Low} & \multicolumn{1}{c|}{{\color[HTML]{010066} High +}}                                                           & {\color[HTML]{FE0000} Low} & \multicolumn{1}{c|}{{Med +}}  & {\color[HTML]{FE0000} Low}     & \multicolumn{1}{c|}{{\color[HTML]{009901} Low}}    & {\color[HTML]{FE0000} Low}            & \multicolumn{1}{c|}{{\color[HTML]{010066} High +}}           & {\color[HTML]{FE0000} Low}                       & {\color[HTML]{009901} Low}                       \\
D                           & {\color[HTML]{FE0000} Low} & \multicolumn{1}{c|}{{\color[HTML]{010066} High +}} & {\color[HTML]{FE0000} Low} & \multicolumn{1}{c|}{{\color[HTML]{009901} Low}}                                                              & {\color[HTML]{FE0000} Low} & \multicolumn{1}{c|}{{\begin{tabular}[c]{@{}c@{}}Med + (NH)\\ Med -- (SH)\end{tabular}}} & {\color[HTML]{FE0000} Low} & \multicolumn{1}{c|}{{Med +}}  & {\color[HTML]{FE0000} Low}     & \multicolumn{1}{c|}{{\color[HTML]{009901} Low}}    & {\color[HTML]{FE0000} Low}            & \multicolumn{1}{c|}{{Med +}}            & {\color[HTML]{FE0000} Low}                       & {\color[HTML]{009901} Low}                       \\
E                           & {\color[HTML]{FE0000} Low} & \multicolumn{1}{c|}{{\color[HTML]{010066} High +}} & {\color[HTML]{FE0000} Low} & \multicolumn{1}{c|}{{\color[HTML]{009901} Low}}                                                              & {\color[HTML]{FE0000} Low} & \multicolumn{1}{c|}{{\color[HTML]{010066} High +}}                                                           & {\color[HTML]{FE0000} Low} & \multicolumn{1}{c|}{{Med --}}  & {\color[HTML]{FE0000} Low}     & \multicolumn{1}{c|}{{\color[HTML]{009901} Low}}    & {\color[HTML]{FE0000} Low}            & \multicolumn{1}{c|}{{\color[HTML]{010066} High +}}       & {\color[HTML]{FE0000} Low}                       & {\color[HTML]{009901} Low}                       \\
F                           & {\color[HTML]{FE0000} Low} & \multicolumn{1}{c|}{{\color[HTML]{010066} High +}} & {\color[HTML]{FE0000} Low} & \multicolumn{1}{c|}{{\color[HTML]{009901} Low}}                                                              & {\color[HTML]{FE0000} Low} & \multicolumn{1}{c|}{{Med --}}                                                           & {\color[HTML]{FE0000} Low} & \multicolumn{1}{c|}{{Med +}}  & {\color[HTML]{FE0000} Low}     & \multicolumn{1}{c|}{{\color[HTML]{009901} Low}}    & {\color[HTML]{FE0000} Low}            & \multicolumn{1}{c|}{{Med +}}            & {\color[HTML]{FE0000} Low}                       & {\color[HTML]{009901} Low}                      
\end{tabular}
\caption{General trends in the variance and relevance of SHAP values for each regime and each feature, where NH refers to the values in the Northern Hemisphere and SH to those in the Southern Hemisphere. To allow direct comparison with LRP, for each regime, we only consider the SHAP values in the region of the regime rather than the whole domain. Therefore + means the feature is actively helpful and -- that it is actively unhelpful.}\label{table:shap_summary}
\end{table}

Figure \ref{SHAP_fig_1} shows the gridpoints for which the sign of the SHAP value remains the same between the 25\% and 75\% quantiles of the ensemble. Note that even though our BNN uses a gridpoint-by-gridpoint approach, for ease of interpretation, we display the SHAP results using a spatial representation, as if SHAP had been applied to a full image. For simplicity, we focus here on Figure \ref{SHAP_fig_1}a which shows the SHAP values for Regime A, although note that the following statements hold true for the regimes for the other figures too. In Figure \ref{SHAP_fig_1}a, red indicates that the probability of Regime A is increased here by including this feature, blue that the probability is decreased and white mainly that this feature has no effect on the probability of predicting Regime A here (although it can also mean there is uncertainty in the SHAP value). If the red matches with the region where the BNN predicts Regime A or the blue matches with the region where the BNN does not predict Regime A, this means that including this feature is actively helpful for predicting this regime in this location. An example of this in Figure \ref{SHAP_fig_1}a is the SHAP values for the longitudinal gradient of the sea level. If, however, the red matches with a region where the BNN does not predict Regime A or the blue matches with the region where the BNN does predict Regime A, then including this feature is actively unhelpful for predicting this regime. An example of this in Figure \ref{SHAP_fig_1}a is the dynamic sea level where including it increases the probability of Regime A everywhere below $40^{\circ}$S and above the North Atlantic Drift, but Regime A is only predicted in certain parts of this region. Notably, Figure \ref{fig:entropy_spatial} shows that at the latitudes where the dynamic sea level is unhelpful, the BNN predictions have high entropy (\textit{i.e.} high uncertainty) suggesting that the dynamic sea level may be a key contributing factor to the uncertainty here. 

\begin{sidewaysfigure}
\begin{subfigure}{\textwidth}
    \centering
    \includegraphics[width=0.85\textwidth]{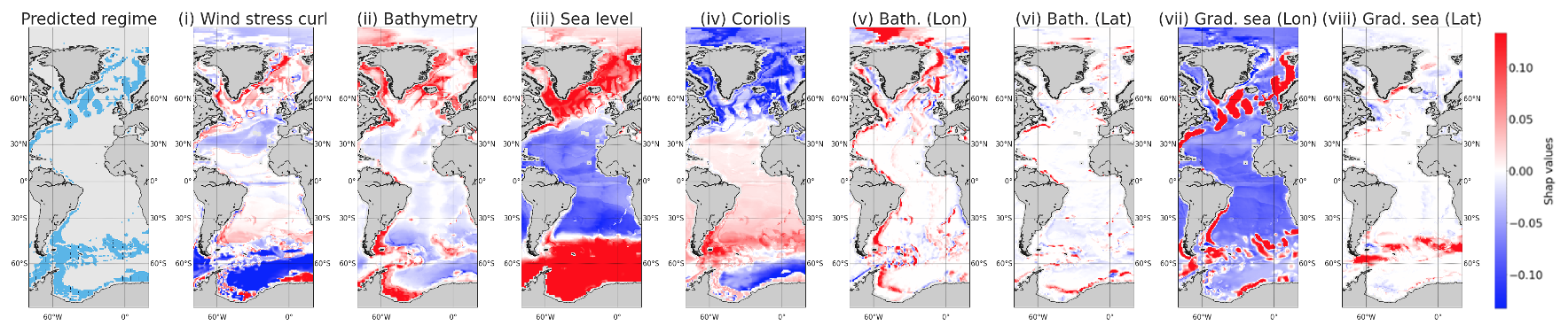}
    \caption{Regime A}
    \label{fig:shap_reg_A}
\end{subfigure}
\begin{subfigure}{\textwidth}
    \centering
    \includegraphics[width=0.85\textwidth]{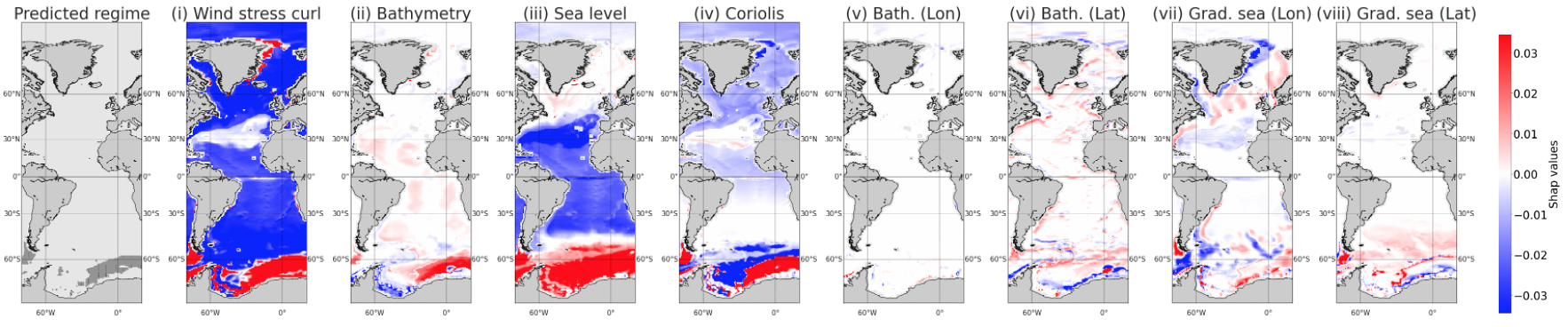}
    \caption{Regime B}
    \label{fig:shap_reg_B}
\end{subfigure}
\begin{subfigure}{\textwidth}
    \centering
    \includegraphics[width=0.85\textwidth]{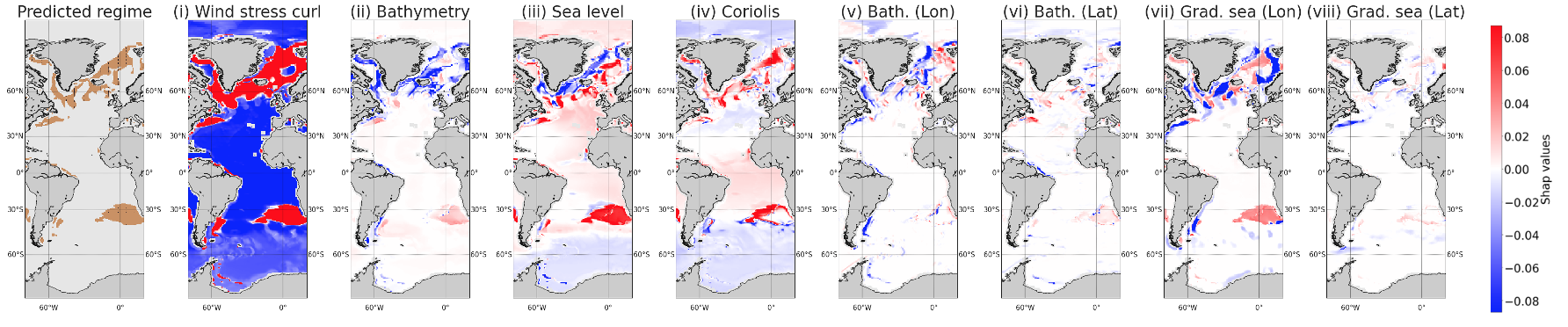}
    \caption{Regime C}
    \label{fig:shap_reg_C}
\end{subfigure}
\caption{SHAP values which are consistent across the whole ensemble for Regimes A (a), B (b), C (c), D (d), E (e) and F (f). Red indicates that the probability of the Regime here is increased by including this feature and blue that the probability is decreased. White means that the SHAP value is either too uncertain or that the variable has no effect.}\label{SHAP_fig_1}
\end{sidewaysfigure}

\begin{sidewaysfigure}
\ContinuedFloat
\begin{subfigure}{\textwidth}
    \centering
    \includegraphics[width=0.85\textwidth]{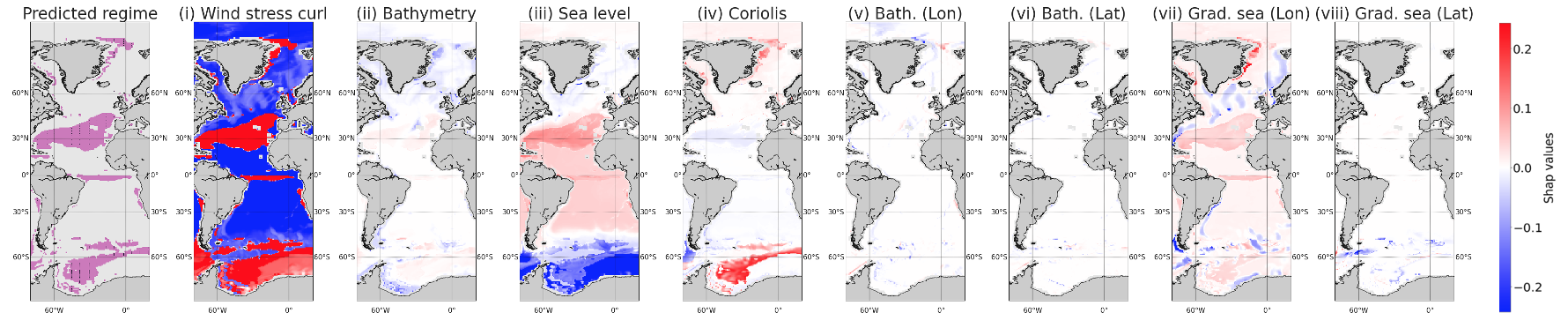}
    \caption{Regime D}
    \label{fig:shap_reg_D}
\end{subfigure}
\begin{subfigure}{\textwidth}
    \centering
    \includegraphics[width=0.85\textwidth]{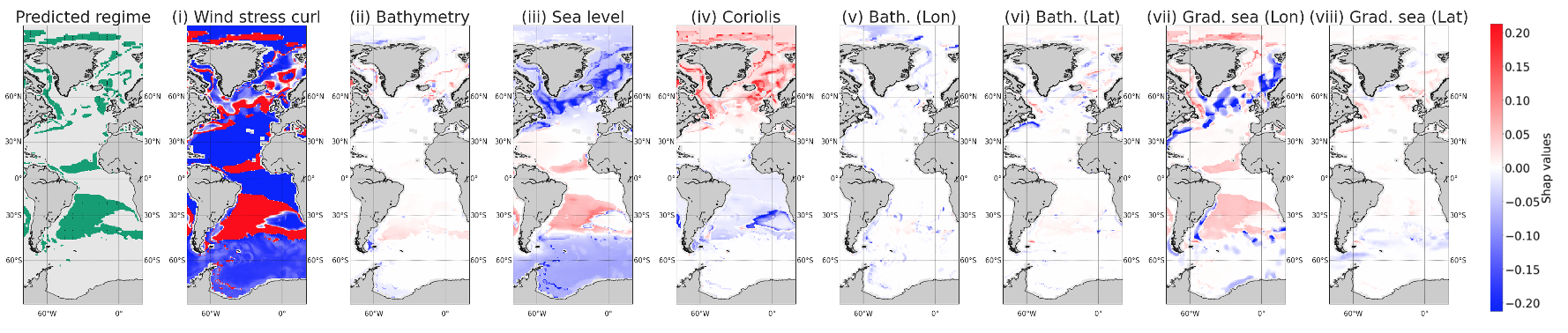}
    \caption{Regime E}
    \label{fig:shap_reg_E}
\end{subfigure}
\begin{subfigure}{\textwidth}
    \centering
    \includegraphics[width=0.85\textwidth]{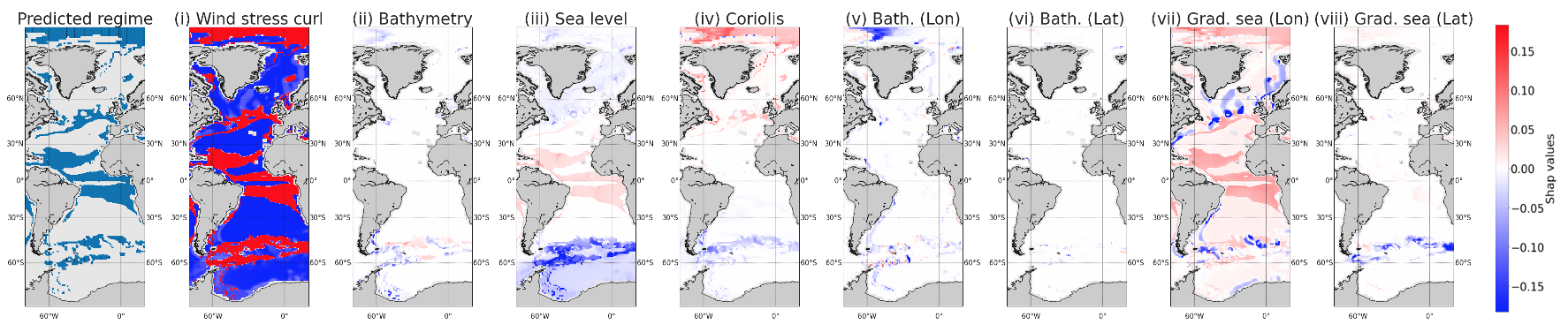}
    \caption{Regime F}
    \label{fig:shap_reg_F}
\end{subfigure}
\caption{SHAP values which are consistent across the whole ensemble for Regimes A (a), B (b), C (c), D (d), E (e) and F (f). Red indicates that the probability of the Regime here is increased by including this feature and blue that the probability is decreased. White means that the SHAP value is either too uncertain or that the variable has no effect.}\label{SHAP_fig_2}
\end{sidewaysfigure}

\begin{table}
\scriptsize
\begin{tabular}{l|cccccccccccccc}
                            & \multicolumn{14}{c}{\textbf{Features}}\\
\multirow{-2}{*}{\textbf{}} & \multicolumn{2}{c|}{\begin{tabular}[c]{@{}c@{}}Wind stress\\ curl\end{tabular}} & \multicolumn{2}{c|}{Bathymetry}                                                 & \multicolumn{2}{c|}{\begin{tabular}[c]{@{}c@{}}Dynamic\\ sea level\end{tabular}} & \multicolumn{2}{c|}{Coriolis}                                                  & \multicolumn{2}{c|}{\begin{tabular}[c]{@{}c@{}}Gradient \\ bathymetry\end{tabular}} & \multicolumn{2}{c|}{\begin{tabular}[c]{@{}c@{}}Gradient \\ sea level\\ (lon)\end{tabular}} & \multicolumn{2}{c}{\begin{tabular}[c]{@{}c@{}}Gradient \\ sea level\\ (lat)\end{tabular}} \\ \cline{2-15} 
                            & Var                        & \multicolumn{1}{c|}{Rel.}                          & Var                        & \multicolumn{1}{c|}{Rel.}                          & Var                         & \multicolumn{1}{c|}{Rel.}                          & Var                        & \multicolumn{1}{c|}{Rel.}                         & Var                          & \multicolumn{1}{c|}{Rel.}                            & Var                                  & \multicolumn{1}{c|}{Rel.}                                     & Var                                             & Rel.                                              \\ \hline
NAD                         & {\color[HTML]{FE0000} Low} & \multicolumn{1}{c|}{{\color[HTML]{010066} High +}} & {\color[HTML]{FE0000} Low} & \multicolumn{1}{c|}{{\color[HTML]{009901} Low}}    & {\color[HTML]{FE0000} Low}  & \multicolumn{1}{c|}{{Med +}}  & {\color[HTML]{FE0000} Low} & \multicolumn{1}{c|}{{Med +}} & {\color[HTML]{FE0000} Low}   & \multicolumn{1}{c|}{{\color[HTML]{009901} Low}}      & {\color[HTML]{FE0000} Low}           & \multicolumn{1}{c|}{{Med +}}             & {\color[HTML]{FE0000} Low}                      & {\color[HTML]{009901} Low}                        \\
GS                          & {\color[HTML]{FE0000} Low} & \multicolumn{1}{c|}{{\color[HTML]{010066} High +}} & {\color[HTML]{FE0000} Low} & \multicolumn{1}{c|}{{Med --}} & {\color[HTML]{FE0000} Low}  & \multicolumn{1}{c|}{{Med --}}  & {\color[HTML]{FE0000} Low} & \multicolumn{1}{c|}{{Med +}} & {\color[HTML]{FE0000} Low}   & \multicolumn{1}{c|}{{\color[HTML]{009901} Low}}      & {\color[HTML]{FE0000} Low}           & \multicolumn{1}{c|}{{\color[HTML]{FE0000} High --}}           & {\color[HTML]{FE0000} Low}                      & {Med +}                      \\
Gyre                         & {\color[HTML]{FE0000} Low} & \multicolumn{1}{c|}{{\color[HTML]{010066} High +}} & {\color[HTML]{FE0000} Low} & \multicolumn{1}{c|}{{\color[HTML]{009901} Low}}    & {\color[HTML]{FE0000} Low}  & \multicolumn{1}{c|}{{\color[HTML]{009901} Low}}    & {\color[HTML]{FE0000} Low} & \multicolumn{1}{c|}{{\color[HTML]{009901} Low}}   & {\color[HTML]{FE0000} Low}   & \multicolumn{1}{c|}{{\color[HTML]{009901} Low}}      & {\color[HTML]{FE0000} Low}           & \multicolumn{1}{c|}{{Med +}}             & {\color[HTML]{FE0000} Low}                      & {\color[HTML]{009901} Low}                        \\ \hline
MAR                         & {\color[HTML]{FE0000} Low} & \multicolumn{1}{c|}{{\color[HTML]{010066} High +}} & {Med} & \multicolumn{1}{c|}{{Med--}}  & {\color[HTML]{FE0000} Low}  & \multicolumn{1}{c|}{{\color[HTML]{009901} Low}}    & {\color[HTML]{FE0000} Low}  & \multicolumn{1}{c|}{{\color[HTML]{009901} Low}}  & {Med}   & \multicolumn{1}{c|}{{Med --}}   & {\color[HTML]{FE0000} Low}           & \multicolumn{1}{c|}{{Med +}}             & {Med}                      & {Med +}                     
\end{tabular}
\caption{Variance and relevance of SHAP values for the key dynamical processes of the North Atlantic Drift (NAD); the Gulf Stream leaving the continental shelf (GS), the wind gyre and the key physical feature of the Mid-Atlantic Ridge as it crosses the wind gyre (MAR) (see Figure \ref{fig:feature_map}).}
\end{table}

As in the LRP section, we also consider the key dynamical processes of the North Atlantic Drift, the Gulf Stream leaving the continental shelf and the North Atlantic wind gyre, as well as the physical characteristic of the mid-Atlantic ridge where it crosses the wind gyre (see Figure \ref{fig:feature_map}). For the North Atlantic Drift, the SHAP values show that the wind stress curl is strongly helpful, and that the Coriolis, dynamic sea level and the longitudinal gradient of the sea level are also helpful. The North Atlantic Drift is a geostrophic current and therefore this feature relevance agrees strongly with the physical theory which governs these types of currents \citep{webb2021introduction}. It is also in contrast to the conclusions from the LRP values where no feature is strongly helpful, only the dynamic sea level and the wind stress are at all helpful and the Coriolis is strongly unhelpful. This difference in the relevance of the Coriolis is also seen for the gyre, which SHAP values say is irrelevant and the LRP values say is strongly unhelpful. Neither agree with intuition from physical theory, which suggests that Coriolis should have some relevance for the gyre. The SHAP values and LRP values do however both identify that for the gyre, the wind stress curl is strongly helpful and the longitudinal gradient of the sea level is helpful, which we recall from Section \ref{LRP_results} agrees with physical intuition. The SHAP and LRP relevance patterns for where the Gulf Stream leaves the continental shelf are also similar to each other. Furthermore, the increased certainty in the SHAP values makes it clear that the longitudinal gradient of the sea level is strongly unhelpful for predictions of this process, whereas for LRP the relevance is very uncertain. Like with LRP, there is also a clear distinction in the SHAP values between the North Atlantic Drift, the Gulf Stream leaving the continental shelf and the wind gyre, strengthening the hypothesis that this feature is an indicator between the three regimes. Finally, the mid-Atlantic ridge is not as prominent in the SHAP values as it is in the LRP values, but the SHAP values still have increased uncertainty there, which is particular significant when the general uncertainty in the ensemble of SHAP values is so low. Furthermore, like the LRP values, the SHAP values also show that both bathymetry and its gradients are more unhelpful at the mid-Atlantic ridge than for the surrounding gridpoints. This supports the conclusions made in Section \ref{LRP_results} that the BNN is able to identify the ridge but not weight it properly. 

To summarise, we have shown that SHAP values provide further evidence of the BNN's ability to identify known physical processes. We have also begun to demonstrate the benefit of using two different XAI techniques, and in the next section compare the findings from the two different techniques more systematically.
\subsubsection{LRP vs. SHAP}\label{sec:shap_vs_LRP}
 As discussed in \ref{sec:shap_method}, LRP and SHAP use two very different approaches to explain skill and hence different types of uncertainty are reflected in their values: LRP considers the neural network parameters and therefore captures the model uncertainty, whereas SHAP captures the sensitivities of the outputs as a result of the uncertainties. Comparing Tables \ref{table:LRP_summary} and \ref{table:shap_summary} clearly shows that this different approach results in SHAP values being more certain in their assessment of feature relevance than LRP values. This difference suggest that our BNN is fairly robust because the uncertainty in the network is greater than the uncertainty in the predictions. This is equivalent to the findings in Section \ref{sec:BNN} where our BNN predictions have low entropy (\textit{i.e.} low uncertainty) despite the weights in the BNN being distributions (see Figure  \ref{fig:entropy_spatial}).

\begin{table}
\tiny
\centering
\begin{tabular}{l|ccccccc}
                            & \multicolumn{7}{c}{\scriptsize{\textbf{Features}}} \\
\multirow{-2}{*}{\textbf{}} & \multicolumn{1}{c|}{\scriptsize{\begin{tabular}[c]{@{}c@{}}Wind stress\\ curl\end{tabular}}}                                  & \multicolumn{1}{c|}{\scriptsize{Bathymetry}}                                                                                 & \multicolumn{1}{c|}{\scriptsize{\begin{tabular}[c]{@{}c@{}}Dynamic\\ sea level\end{tabular}}}                              & \multicolumn{1}{c|}{\scriptsize{Coriolis}}                                                                                     & \multicolumn{1}{c|}{\scriptsize{\begin{tabular}[c]{@{}c@{}} Gradient \\ \hspace{2pt} bathymetry \hspace{2pt}\end{tabular}}} & \multicolumn{1}{c|}{\scriptsize{\begin{tabular}[c]{@{}c@{}}Gradient \\ sea level\\ (lon)\end{tabular}}}                        & \scriptsize{\begin{tabular}[c]{@{}c@{}}Gradient \\ sea level\\ (lat)\end{tabular}}                    \\ \hline
A                           & \multicolumn{1}{c|}{{\color[HTML]{FE0000}  \hspace{2pt} Med -- \textgreater Med +}}           & \multicolumn{1}{c|}{{ \hspace{2pt} Med + \textgreater High +}} & \multicolumn{1}{c|}{{=}}                                    & \multicolumn{1}{c|}{{=}}                           & \multicolumn{1}{c|}{{=}}                                       & \multicolumn{1}{c|}{{\color[HTML]{FE0000} \hspace{2pt} High -- \textgreater High +}}                    & {Med -- \textgreater Low}                        \\
B                           & \multicolumn{1}{c|}{{=}}                                   & \multicolumn{1}{c|}{{\color[HTML]{FE0000} \hspace{2pt} Med -- \textgreater Med +}} & \multicolumn{1}{c|}{{=}}                                    & \multicolumn{1}{c|}{{Low \textgreater High +}}    & \multicolumn{1}{c|}{{=}}                                       & \multicolumn{1}{c|}{{=}}                                              & {=}                                              \\
C                           & \multicolumn{1}{c|}{{=}}                                   & \multicolumn{1}{c|}{{=}}                         & \multicolumn{1}{c|}{{\hspace{2pt} Low \textgreater High +}}              & \multicolumn{1}{c|}{{Low \textgreater Med +}}      & \multicolumn{1}{c|}{{=}}                                       & \multicolumn{1}{c|}{{Low \textgreater Med +}}                         & {=}                                              \\
D                           & \multicolumn{1}{c|}{{=}}                                   & \multicolumn{1}{c|}{{Med -- \textgreater Low}}   & \multicolumn{1}{c|}{{=}}                                    & \multicolumn{1}{c|}{{\color[HTML]{FE0000} \hspace{2pt} Med -- \textgreater Med +}}   & \multicolumn{1}{c|}{{=}}                                       & \multicolumn{1}{c|}{{=}}                                              & {=}                                              \\
E                           & \multicolumn{1}{c|}{{=}}                                   & \multicolumn{1}{c|}{{=}}                         & \multicolumn{1}{c|}{{=}}                                    & \multicolumn{1}{c|}{{\hspace{2pt} High -- \textgreater Med --}} & \multicolumn{1}{c|}{{=}}                                       & \multicolumn{1}{c|}{{=}}                                              & {Med + \textgreater Low}                         \\
F                           & \multicolumn{1}{c|}{{\color[HTML]{FE0000} \hspace{2pt} Med -- \textgreater High +}}          & \multicolumn{1}{c|}{{Med -- \textgreater Low}}   & \multicolumn{1}{c|}{{=}}                                    & \multicolumn{1}{c|}{{\color[HTML]{FE0000} \hspace{2pt} Med -- \textgreater Med +}}   & \multicolumn{1}{c|}{{=}}                                       & \multicolumn{1}{c|}{{=}}                                              & {Med \textgreater Low} 
\end{tabular}
\caption{Comparing the general trend in the relevances of LRP \textgreater SHAP. If the relevance changes sign, the change is coloured red.}\label{table:comparison}
\end{table}

Table \ref{table:comparison} directly compares the trends in the relevances of LRP and SHAP. Some differences between SHAP and LRP are due to the fact that SHAP values separate out the relevance of each feature for each regime, whereas LRP values consider the relevance of a feature for all regimes simultaneously. For example, in the upper part of the Atlantic ($\sim 60^{\circ}$N), the SHAP values for Regime A (Figure \ref{SHAP_fig_1}a) show that the wind stress curl is helpful for predicting that regime. However, the SHAP values for regimes C and  E (Figures \ref{SHAP_fig_1}c and \ref{SHAP_fig_2}e respectively) show that the wind stress curl also increases the probability of regimes C and E at that location. Therefore when the SHAP values for all regimes are considered, the wind stress curl may actually be more unhelpful than helpful, agreeing with LRP.

As in Sections \ref{LRP_results} and \ref{sec:shap_results}, for brevity we do not discuss all differences between SHAP and LRP. Instead, we summarise the key comparisons for each regime in the following list:

\vspace{10pt}

\textbf{Regime A}
\begin{itemize}
    \item Wind stress curl is helpful in SHAP but unhelpful in LRP (see discussion in text previously).
    \item The locations where the dynamic sea level has strong relevance in the LRP values coincides directly with the regions where regime A is predicted. The dynamic sea level is also helpful in SHAP, but SHAP shows that this feature also increases the probability of Regime A in areas where Regime A is not predicted. Note that the latter are areas of high entropy (see Figure \ref{fig:entropy_spatial}).
    \item The longitudinal gradient of the dynamic sea level is strongly unhelpful in LRP and strongly helpful in SHAP. Again this region of difference corresponds to areas of high entropy in the BNN predictions.
\end{itemize}

\textbf{Regime B}
\begin{itemize}
    \item Wind stress curl is strongly helpful in both LRP and SHAP, but along the east coast of Greenland, in the SHAP values, the wind stress curl increases the probability of regime B, but the BNN does not predict this regime nor would regime B be accurate there. This region has high entropy and in the LRP values the relevance of the wind stress curl switches here from unhelpful in the 25th quantile to helpful in the 75th quantile. This suggests that the BNN has high uncertainty in the relevance of this input feature here.
    \item In the SHAP values, the bathymetry is helpful but in LRP it is unhelpful. This is despite the fact that regions where this regime is predicted by the BNN, generally have low entropy
    \item Coriolis is strongly helpful in SHAP (as would be expected from physical intuition) but has low relevance in the LRP values, apart from around the tip of South America where it is strongly helpful.
\end{itemize}

\textbf{Regime C}
\begin{itemize}
    \item In regime C, particularly in the southern hemisphere, most features have no relevance in the LRP values but a medium or high relevance in the SHAP values. In particular, the dynamic sea level and its longitudinal gradient have no relevance with high certainty in the LRP values but strong positive relevance with high certainty in the SHAP values. Note that entropy is low for this regime, particularly in the southern hemisphere 
    \item Wind stress curl is strongly helpful in both LRP and SHAP. This likely explains the irrelevance in other features in the LRP values: LRP values consider the weightings in the BNN, and the wind stress curl has such a strong weighting that all other features are comparatively close to zero. In contrast, SHAP values consider the sensitivity of the output to other features, which does change
\end{itemize}

\textbf{Regime D}
\begin{itemize}
    \item In both SHAP and LRP, the dynamic sea level is helpful in the northern hemisphere but unhelpful in the southern hemisphere.
    \item Coriolis is strongly helpful at high latitudes in the SHAP values and irrelevant at mid-latitudes. In contrast, Coriolis is unhelpful in the LRP values especially at the mid-latitudes.  This variation suggests the BNN does not accurately weight low values of Coriolis (near the equator), resulting in unhelpful LRP values. Nearer the poles, the weighting improves enough for SHAP to become helpful but not enough for LRP to become helpful.
    \item The wind stress curl is strongly helpful in both the SHAP and LRP values but the SHAP values for wind stress curl do not have increased uncertainty at the mid-atlantic ridge. This reflects the general trend of greater certainty in SHAP values than LRP values.
\end{itemize}

\textbf{Regime E}
\begin{itemize}
    \item Wind stress curl is strongly helpful for SHAP and LRP, but the LRP values in the southern hemisphere have high variance especially around $35^{\circ}$S where the BNN entropy is highest.
    \item Coriolis is strongly unhelpful in LRP especially at mid-latitudes but only slightly unhelpful in SHAP (see discussion for Regime D).
    \item The latitudinal gradient of the dynamic sea level is irrelevant in the SHAP values but has relevance in the LRP values. There is however a split in the LRP relevance at $35^{\circ}$S -- above this latitude the relevance is positive and below the relevance is negative. This split corresponds with an increase in entropy, where entropy is higher below this latitude.
\end{itemize}

\textbf{Regime F}
\begin{itemize}
    \item Wind stress curl is strongly helpful in SHAP but unhelpful in LRP. We would expect wind stress curl to be helpful from Table \ref{table:regimes} so this is an example where SHAP agrees more closely with physical intuition than LRP.
    \item Bathymetry is unhelpful for this regime in LRP but in SHAP only has relevance at the coastlines. 
    \item Coriolis is unhelpful in LRP at mid-latitudes but has no relevance in SHAP except at high latitudes (see discussion for Regime D).
    \item The latitudinal gradient of the dynamic sea level is very uncertain in LRP changing from unhelpful to helpful, despite the fact that the entropy is low for predictions of this regime. This gradient has no relevance according to SHAP , and thus the mean of the SHAP and LRP values agree for this feature. This reflects the general trend of greater certainty in SHAP values than LRP values.
\end{itemize}
  
In general, SHAP and LRP agree on how to explain the skill of the BNN, thus meaning that in our work we do not have a `disagreement problem'. There are however some small differences, which can either be explained by the different ways in which these two techniques interpret skill or by the fact that they occur where there is high entropy in the BNN predictions reflecting the BNN's uncertainty in feature relevance. We have thus demonstrated that both techniques are helpful for understanding the BNN's interpretations of physical processes. Moreover, where the two techniques agree with each other and in particular also agree with physical intuition, this greatly improves the trustworthiness of the feature relevance explanations in the BNN and where the techniques differ between themselves and/or with physical intuition there is scope for further analysis and learning of both BNN and physical ocean processes.

\section{Discussion and Conclusion}\label{sec:conclusion}
In this work, we have successfully applied a BNN and two different XAI techniques to explore the trustworthiness of ocean dynamics predictions made using a machine learning technique. We have shown that using a BNN rather than a classical deterministic neural network adds considerable value to predictions, by making uncertainty analysis possible and allowing practitioners to make informed decisions as to whether to trust a prediction or conduct further investigation. Furthermore, our analysis of the entropy (\textit{i.e.} uncertainty) of the BNN predictions shows the promising result that the predictions are notably more certain when they are correct than when they are incorrect. 

Through our novel applications of the XAI techniques, LRP and SHAP, we have also shown that it is possible to explain the skill of a BNN, conduct uncertainty analysis of explainability values, and hence use XAI techniques to understand the extent to which the BNN is fit for purpose, where we here demonstrate this using comparison with theory. Our spatial representation of both the SHAP and LRP values means that the relevance of specific important dynamical processes such as the North Atlantic Drift can be identified, thereby improving the interpretability and hence trustworthiness of the results. This comparison with physical theory is important to ensure that what the BNN has learned is genuinely rooted in physical theory. Moreover, the spatial coherency of both the uncertainty and XAI assessments suggest that our framework could be leveraged to identify potential new physical hypotheses in areas of interest, guided by the BNN's ability to highlight hitherto unrecognised correlations in the input space. However, we stress that these correlations do not necessarily imply causation \citep{samek2021explaining}. Therefore for deployment of developed neural network applications  for high-stakes decision making within geoscience, these correlations should only be used to postulate new hypotheses, which must then be explored using a well-conducted study driven by physical theory. 

Our comparison of LRP and SHAP values has shown that in general they agree with each other as to which features are relevant in each area of the domain, building trust in the BNN predictions and their explanations. This is particularly striking given that SHAP is model-agnostic and does not consider any internal architecture of the network, exploring only how sensitive the predictions are to the removal of input features, whereas LRP uses a model-intrinsic approach based on the internal architecture of the network. These two different XAI techniques do result in different levels of uncertainty in the feature relevances because LRP better captures the neural network model uncertainty and SHAP better captures BNN prediction sensitivity. Any disagreements in feature relevance also tend to occur due to these different approaches and/or in regions of high entropy. Knowledge of these disagreements is useful to practitioners as it highlights areas where the explanation of the BNN's skill is less trustworthy and may require further analysis. Furthermore the use of an ocean dynamical framework allows the accuracy of the XAI results in this work to be verified with physical intuition. It also enables a better understanding of how SHAP and LRP explain skill which is beneficial to the machine learning community. Where there are differences between the XAI techniques and physical intuition, this provides another potential opportunity to learn more about physical theory, although with the same caveats discussed above.

We hypothesise that the good agreement with physical intuition demonstrated in this work is in part due to the overall normally distributed covariance structure of the problem, which is helpful for the K-means clustering and thus directly beneficial for the BNN training \citep{sonnewald2019unsupervised}. The methodology outlined in this work has many potential applications in geoscience and beyond, for more complex and nonlinear covariance structures. Besides classification problems, where the re-application of our methodology is straightforward, a promising research avenue is the use of XAI, augmented with uncertainty quantification, for regression problems. An example of high interest to the climate modeling community is subgrid scale parametrization efforts for numerical models. So far, subgrid scale parametrizations based on neural networks have limited generalization capacities, especially in areas of the numerical model space that they are not explicitly trained on \citep{bolton2019applications}. A regression based XAI framework could thus accelerate the development of such techniques, because the reasons why the networks fail to generalise might be better understood for both specific local scale features such as where the Gulf Stream leaves the continental shelf and larger scale processes. In further work, we will benefit from the ongoing recent research developments in XAI for regression, for example in \cite{letzgus2021toward}, and aim to apply our methodology to this more challenging problem. 

Finally, we recommend that for trustworthy explainability results for more complex covariance structures, a BNN should be used along with one model-intrinsic XAI technique, like LRP and one model-agnostic XAI technique like SHAP, so as to consider both neural network model properties and output sensitivity. For an accurate and robust network, we would expect the similarities between the two XAI techniques to dominate and the differences to highlight areas that require further analysis, thus being of valuable use to practitioners and might hint at new scientific hypotheses. 

\subsubsection*{Data Availability Statement}
	The relevant code for the explainable Bayesian THOR framework presented in this work is preserved at \cite{THOR_code}, available via CC-BY licence. The ECCOv4r3 data is available to download at \cite{ECCO_data}.

\subsubsection*{Acknowledgements}
MCAC acknowledges funding from the Higher Education, Research and Innovation Department at the French Embassy in the United Kingdom. MS acknowledges funding from the Cooperative Institute for Modeling the Earth System, Princeton University, under Award NA18OAR4320123 from the National Oceanic and Atmospheric Administration, U.S. Department of Commerce. The statements, findings, conclusions, and recommendations are those of the authors and do not necessarily reflect the views of Princeton University, the National Oceanic and Atmospheric Administration, or the U.S. Department of Commerce. RL acknowledges the Make Our Planet Great Again (MOPGA) fund from the Agence Nationale de Recherche under the ``Investissements d'avenir'' programme with reference ANR-17-MPGA-0010.

\appendix
\section{LRP figures}\label{appendix}
\begin{figure}
\begin{subfigure}{0.49\textwidth}
    \centering
    \includegraphics[height = 0.5\textwidth]{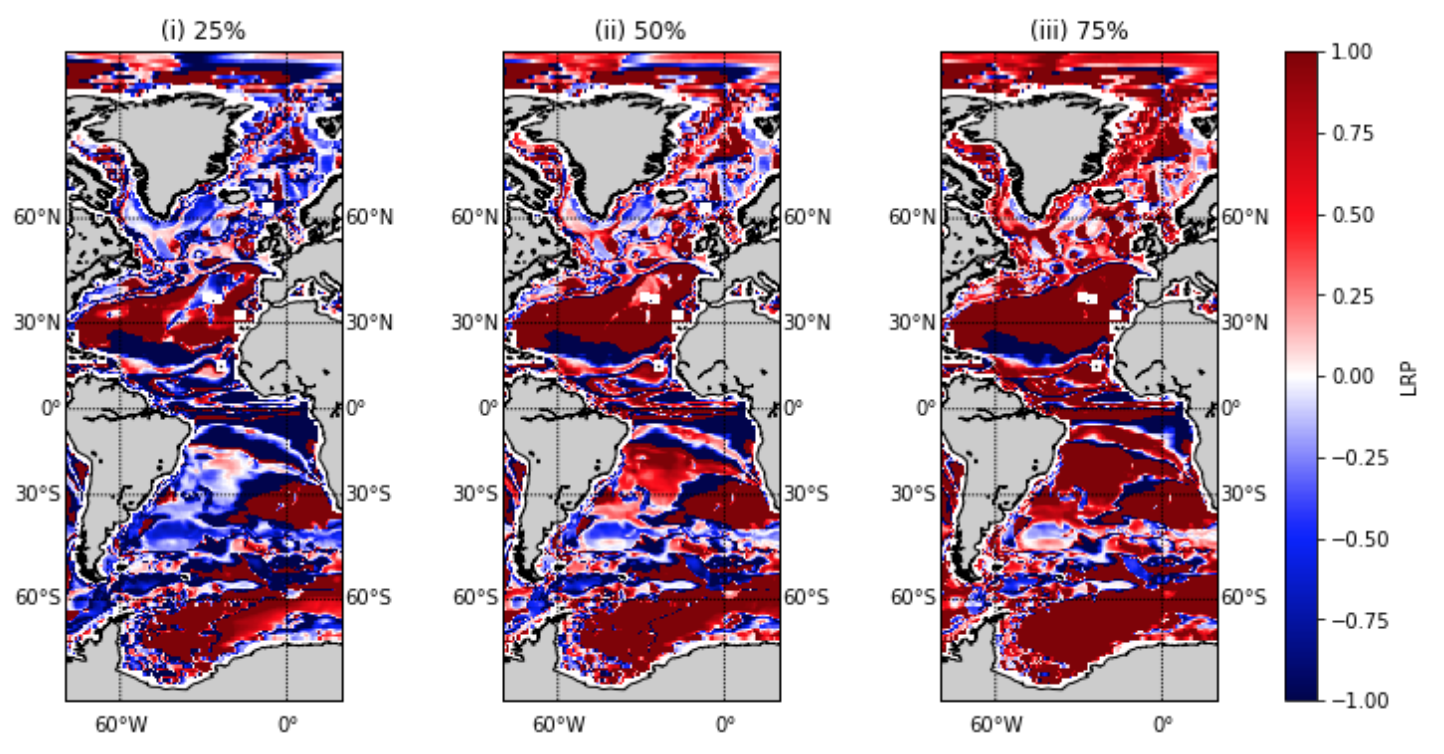}
    \caption{Wind stress curl.}
    \label{fig:quantile_curltau}
\end{subfigure}
\hfill
\begin{subfigure}{0.49\textwidth}
    \centering
    \includegraphics[height = 0.5\textwidth]{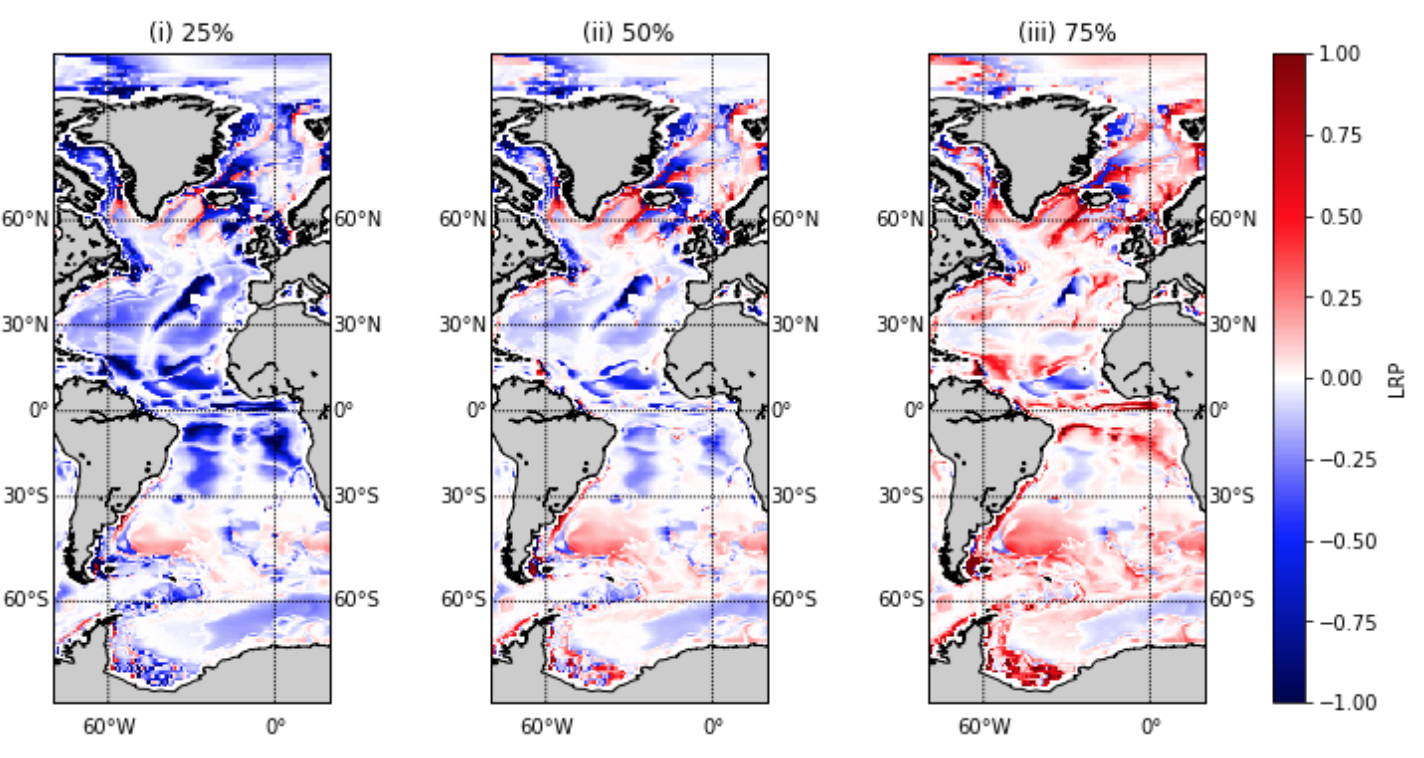}
    \caption{Bathymetry.}
    \label{fig:quantile_bath}
\end{subfigure}
\begin{subfigure}{0.49\textwidth}
    \centering
    \includegraphics[height = 0.5\textwidth]{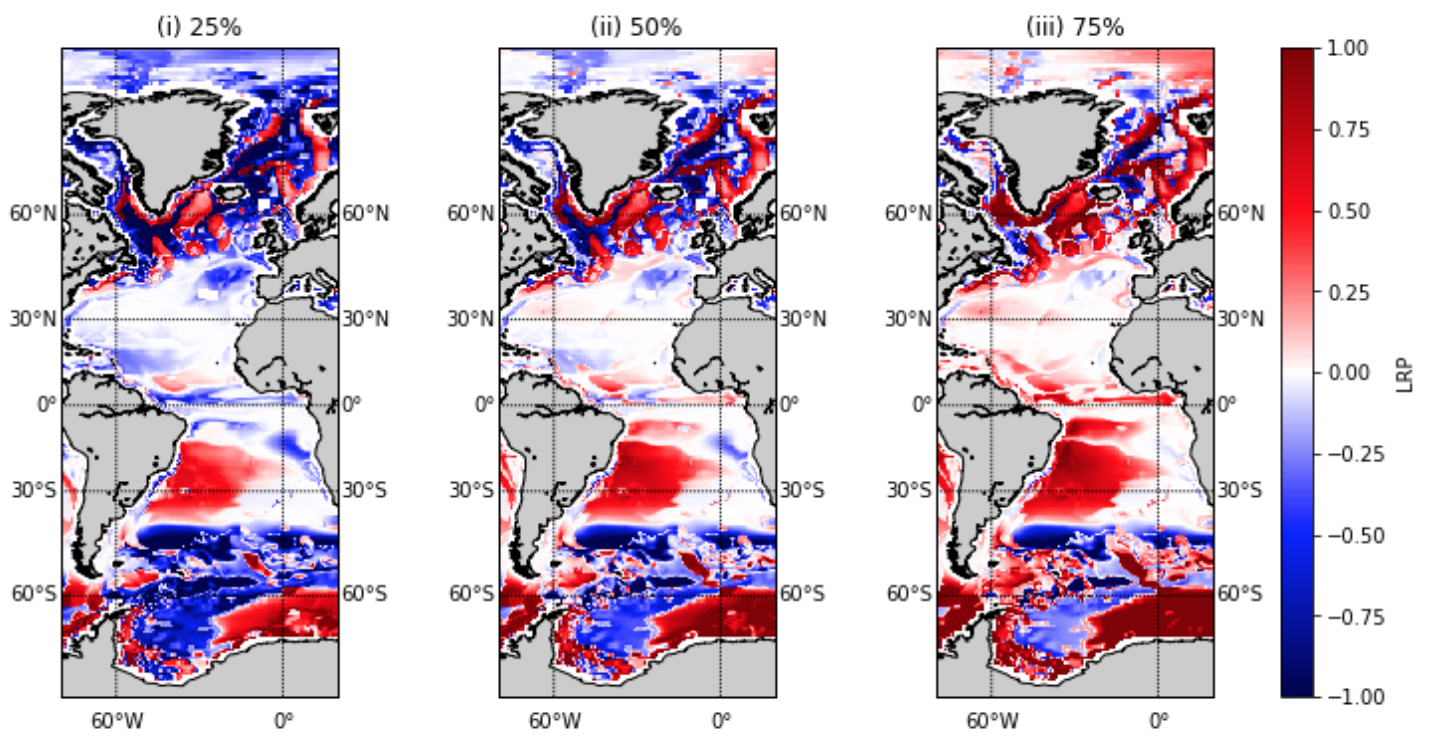}
    \caption{Dynamic sea level.}
    \label{fig:quantile_ssh}
\end{subfigure}
\hfill
\begin{subfigure}{0.49\textwidth}
    \centering
    \includegraphics[height = 0.5\textwidth]{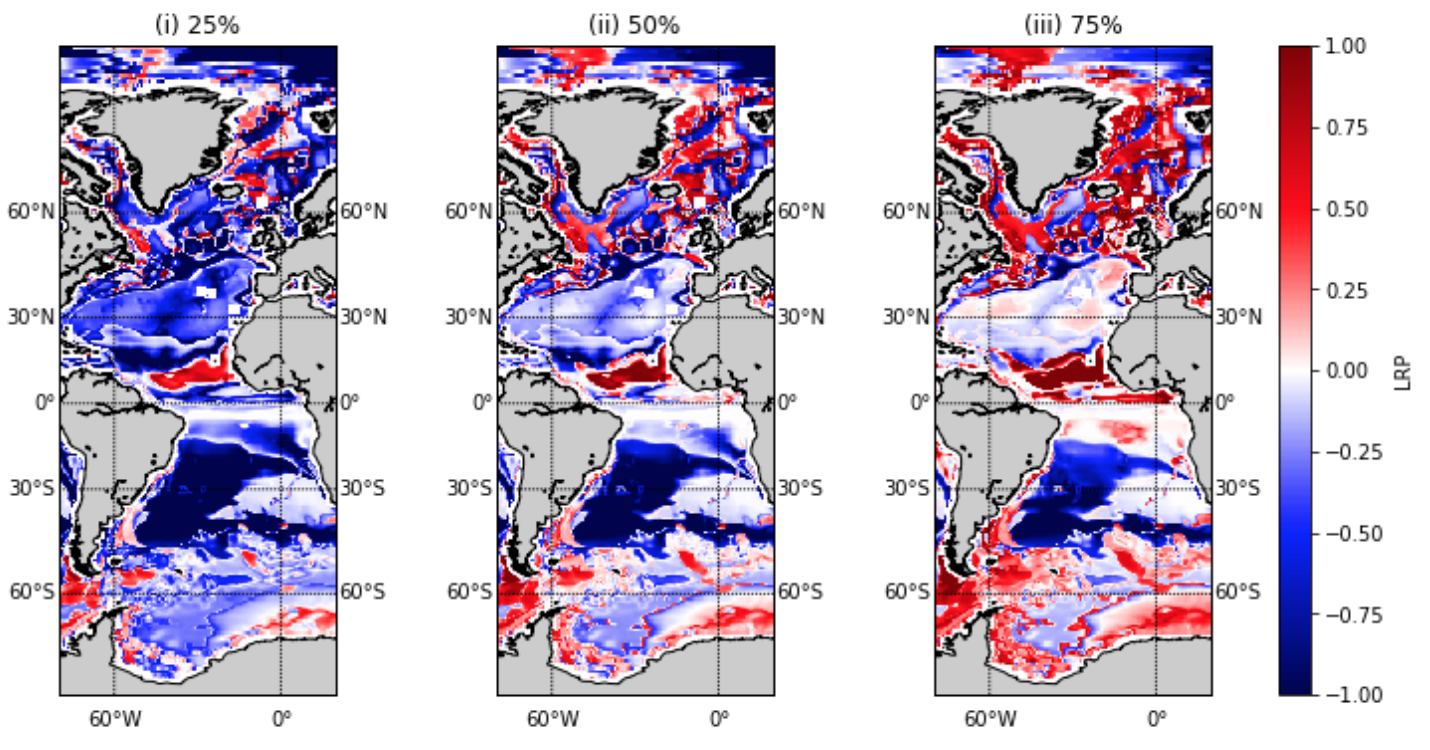}
    \caption{Coriolis force.}
    \label{fig:quantile_F}
\end{subfigure}
\begin{subfigure}{0.49\textwidth}
    \centering
    \includegraphics[height = 0.5\textwidth]{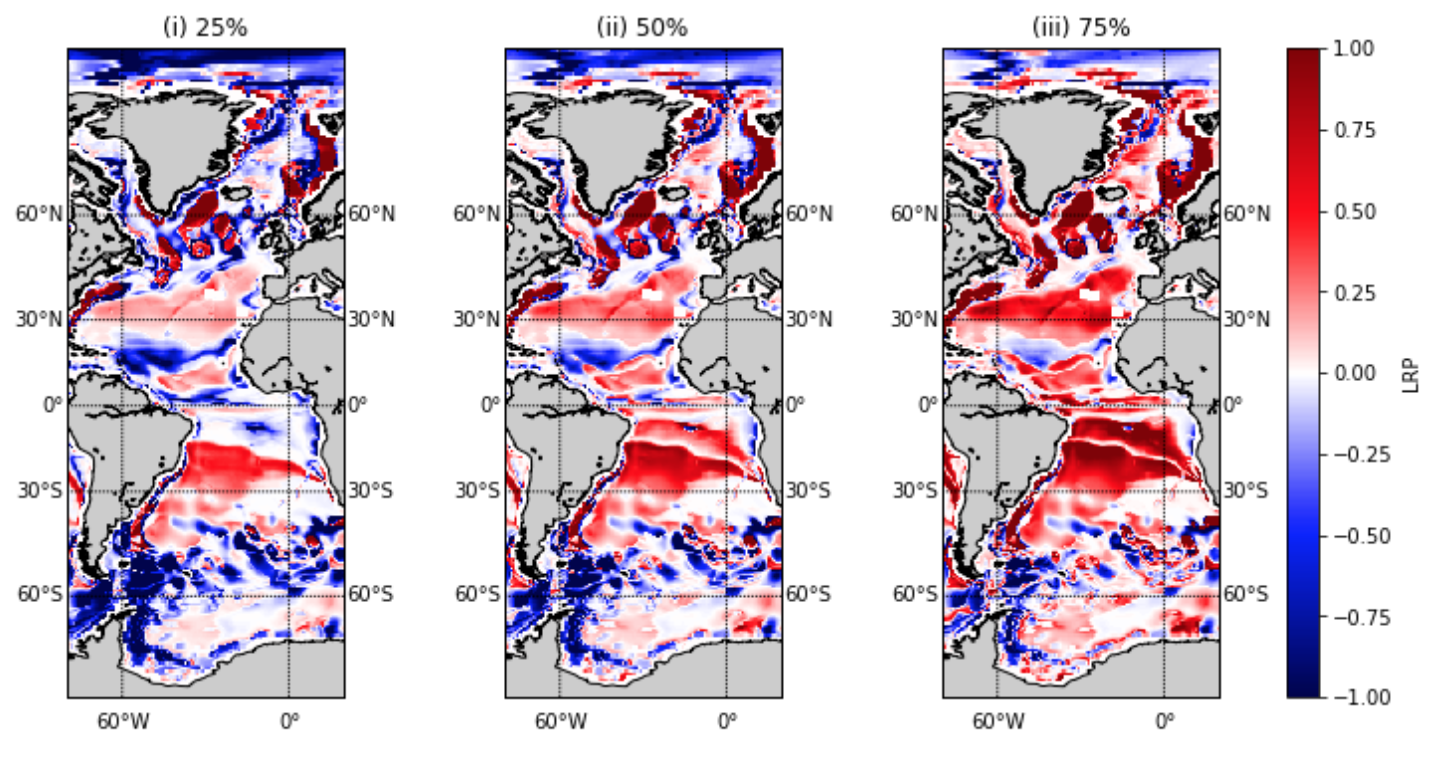}
    \caption{Gradient dynamic sea level (lon).}
    \label{fig:quantile_sshx}
\end{subfigure}
\hfill
\begin{subfigure}{0.49\textwidth}
    \centering
    \includegraphics[height = 0.5\textwidth]{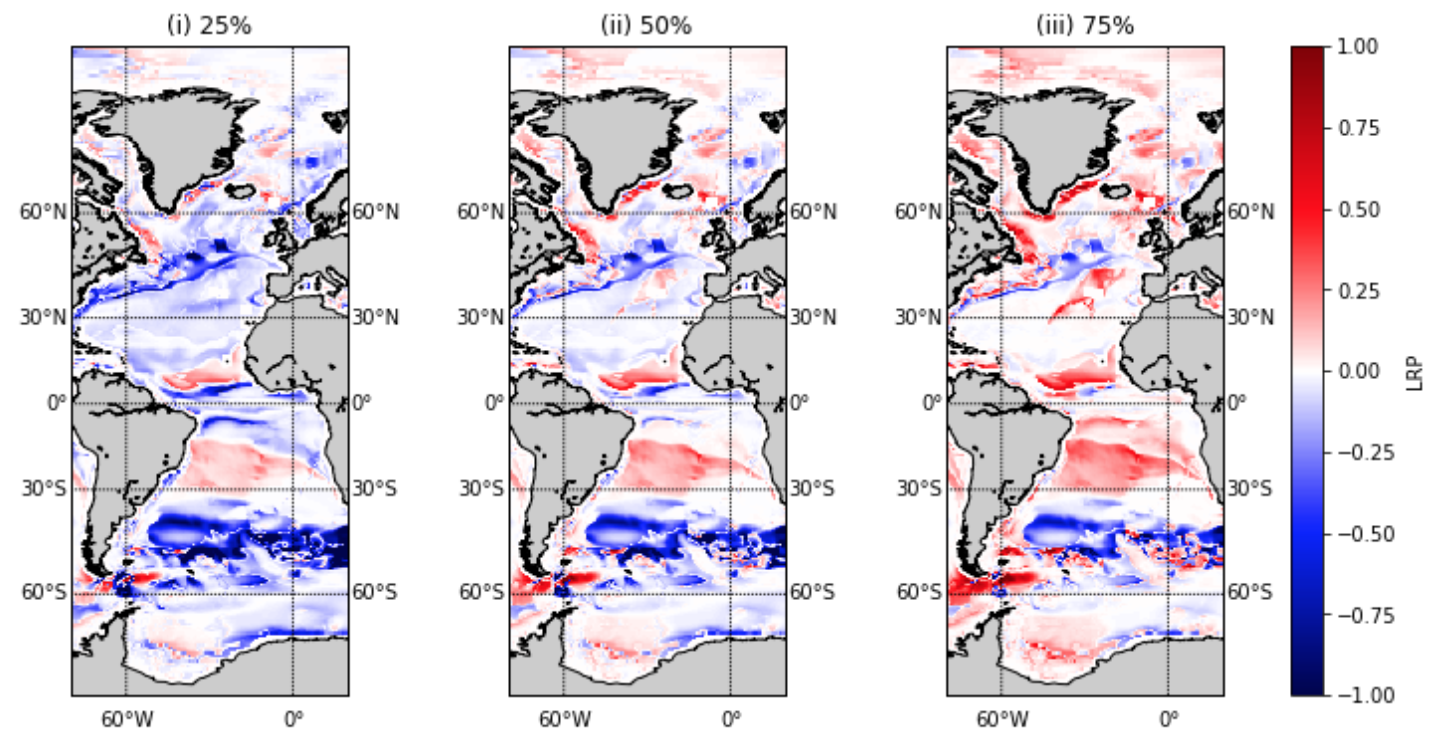}
    \caption{Gradient dynamic sea level (lat).}
    \label{fig:quantile_sshy}
\end{subfigure}

\begin{subfigure}{0.49\textwidth}
    \centering
    \includegraphics[height = 0.5\textwidth]{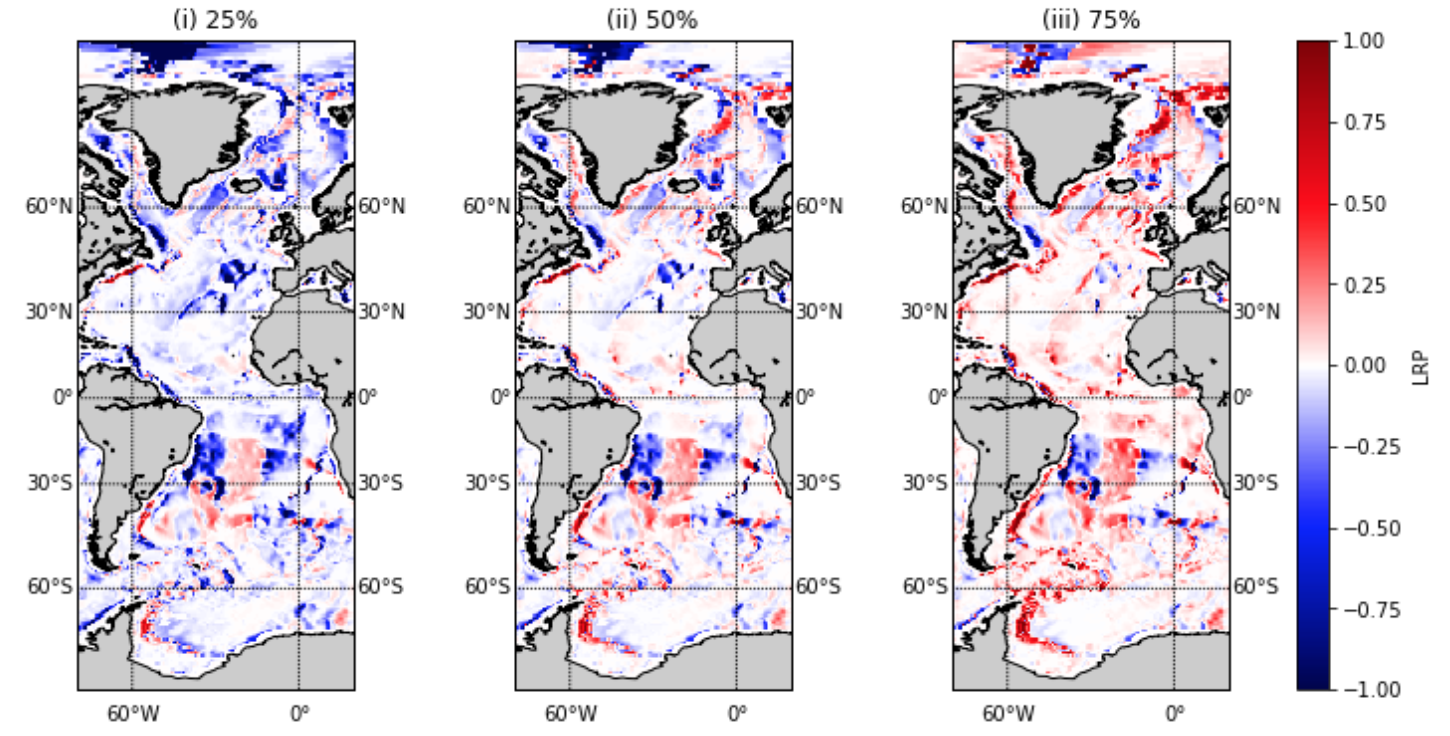}
    \caption{Gradient bathymetry (lon).}
    \label{fig:quantile_bathx}
\end{subfigure}
\hfill
\begin{subfigure}{0.49\textwidth}
    \centering
    \includegraphics[height = 0.5\textwidth]{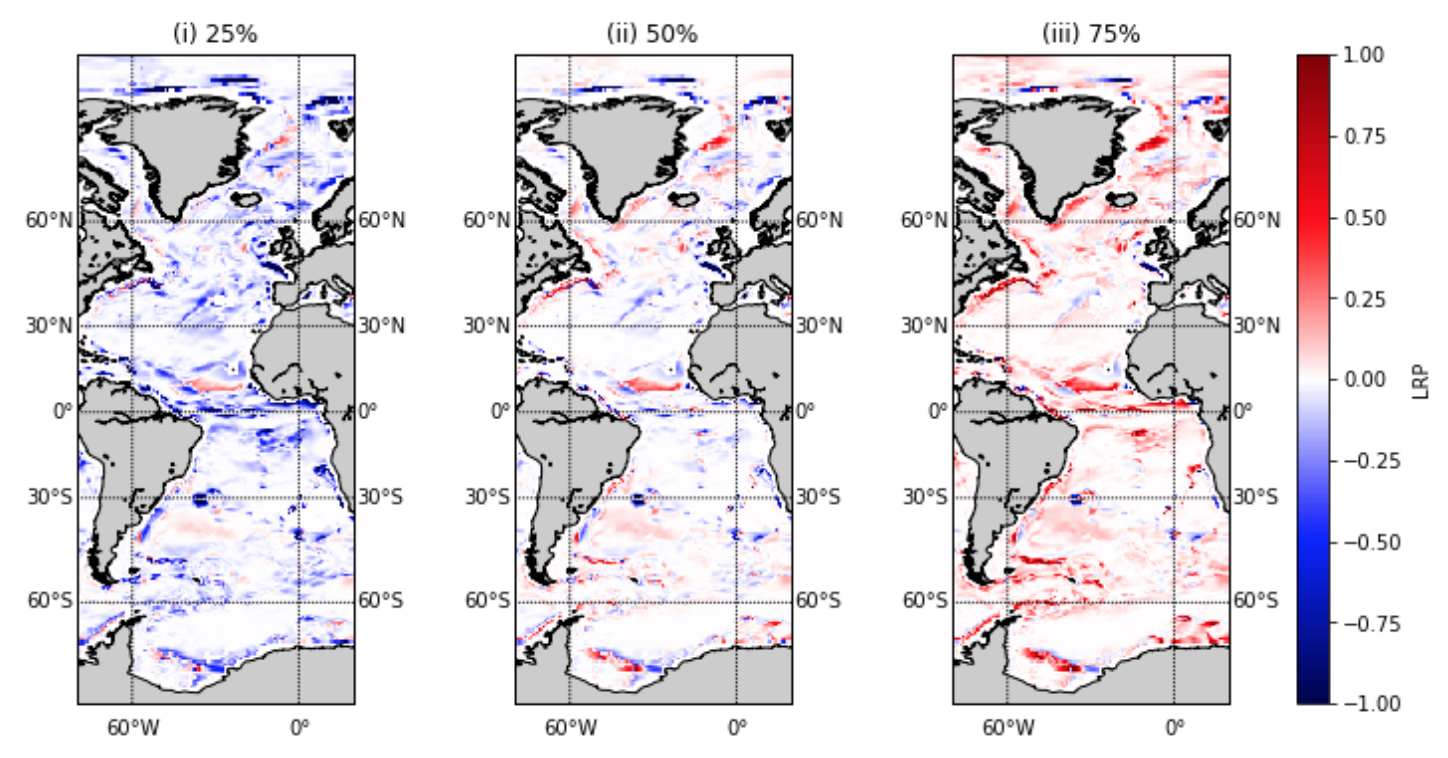}
    \caption{Gradient bathymetry (lat).}
    \label{fig:quantile_bathy}
\end{subfigure}
\caption{LRP values at the 25th, 50th (median) and 75th quantile of the ordered ensemble.}\label{fig:LRP_quantile}
\end{figure}

Figure \ref{LRP_consistent} in Section \ref{LRP_results} reveals the LRP values which have a consistent sign across the 25\%, 50\% and 75\% quantiles. However, there is also considerable variability across the ensemble of LRP values and thus to give a better idea of this uncertainty, we also include Figure \ref{fig:LRP_quantile} which shows the 25\%, 50\% and 75\% quantiles of the LRP ensemble. Using this figure, we see, for example, that for many regions the bathymetry gradients go from being strongly unhelpful at the 25\% quantile to strongly helpful at the 75\% quantile, showing a high degree of uncertainty. The figure also illustrates better the regions which are irrelevant to BNN predictions (\textit{i.e.} where the LRP value is zero).

\bibliographystyle{apalike}
\bibliography{sample}

\end{document}